\documentclass[sigconf,nonacm]{acmart}

\usepackage{amsmath}    
\usepackage{tabularx} 
\usepackage{rotating}
\usepackage{booktabs}
\usepackage{subcaption} 
\usepackage[utf8]{inputenc}
\usepackage{longtable}
\usepackage{csquotes}
\usepackage{multirow}
\usepackage{enumitem}

\AtBeginDocument{%
  }

\begin{document}

\title{Bridging Voting and Deliberation with Algorithms: Field Insights from vTaiwan and Kultur Komitee} 

\author{Joshua C. Yang}
\email{joyang@ethz.ch}
\orcid{0009-0004-4044-6338}
\affiliation{%
  \institution{ETH Zurich}
  \city{Zurich}
  \country{Switzerland}
}

\author{Fynn Bachmann}
\email{fbachmann@ifi.uzh.ch}
\orcid{0000-0001-9571-5671}
\affiliation{%
  \institution{University of Zurich}
  \city{Zurich}
  \country{Switzerland}
}

\begin{abstract} 

Democratic processes increasingly integrate large-scale voting with face-to-face deliberation to reconcile individual preferences with collective decision-making. This work introduces algorithmic methods to bridge online voting with face-to-face deliberation, tested in two real-world scenarios: Kultur Komitee 2024 (KK24) and vTaiwan. We present three key contributions: (1) Preference-based Clustering for Deliberation (PCD), enabling both focused and broad discussions by computing balanced homogeneous and heterogeneous groups; (2) Human-in-the-loop MES, enhancing the Method of Equal Shares algorithm with real-time feedback, giving participants control over algorithmic decision-making; and (3) the ReadTheRoom method, using opinion mapping to identify agreement and divergence while tracking opinion shifts during deliberation. These actionable frameworks extend in-person deliberation with scalable digital methods that address the complexities of modern participatory decision-making.
\footnote{© The definitive Version of Record was published in:\\
    \textit{Proceedings of the 2025 ACM Conference on Fairness, Accountability, and Transparency (FAccT '25), June 23–26, 2025, Athens, Greece.}\\
    DOI: \url{https://doi.org/10.1145/3715275.3732205}
    }
\end{abstract}

\ccsdesc[500]{Human-centered computing~Social computing theory, methods, and systems}
\ccsdesc[300]{Computing methodologies~Voting methods}
\ccsdesc[300]{Information systems~Collaborative and social computing systems and tools}
\ccsdesc[200]{Human-centered computing~Interaction design}

\keywords{Participatory budgeting, citizens' assemblies, Method of Equal Shares, group deliberation, democratic innovations, social computing, voting systems}

\maketitle

\section{Introduction}

In an era where democratic decision-making faces increasing complexity, computational tools have emerged as essential enablers of fair and scalable processes. Modern challenges in democracy often involve balancing individual preferences with collective decisions \cite{weyl2024}, particularly in large-scale contexts. As \citet{hendriks2024} point out, each democratic format offers unique strengths and limitations. Deliberative designs are effective in building meaningful discussions, but often engage a limited number of participants and struggle to achieve large-scale impact \cite{michels2011}. Conversely, plebiscitary designs, or voting, can involve broad citizen participation and produce decisive collective outcomes, but often lack the depth of deliberative dialogue. To address these trade-offs, hybrid democratic innovations have emerged, combining the strengths of both approaches \cite{felicetti2021,hendriks2023}.

Deliberative democracy has long focused on the role of informed discussions in determining collective decisions~\cite{bachtiger_deliberative_2018}. Recent studies, however, emphasise a complementary dynamic: voting can enhance deliberation~\cite{bachtiger_deliberation_2018-3,bachtiger_democratic_2018}. Namely, voting complements and enhances deliberation in seven key ways, as pointed out by \citet{chambers2023}: it (i) provides a feasible and fair closure mechanism; (ii) ensures equal recognition and status among participants; (iii) politicises deliberation by internalising conflict; (iv) induces authenticity by encouraging participants to reveal their preferences; (v) preserves dissent; (vi) defines issues to focus deliberation effectively; and, (vii) in public voting contexts, fosters accountability for claims. Understanding these dynamics helps identify designs that better integrate deliberation and voting to capitalise on their combined strengths. 

Fair mechanisms for aggregating diverse preferences are essential for proportional representation in collective decision-making, ensuring outcomes reflect the priorities of all groups rather than only favouring majorities \cite{aziz2020}. Algorithms like the Method of Equal Shares (MES) \cite{peters2021} enable fair resource allocation while protecting minority voices. However, real-world implementation faces challenges, including bridging technological literacy gaps, managing real-time deliberations, and building trust in algorithmic outcomes. To address these issues, accessible and user-friendly tools are vital for translating theoretical fairness into practical, inclusive applications \cite{yang2024}. 

This research offers practical insights into designing democratic innovations that integrate voting and deliberation. In particular, we introduce three computational methods applied in two real-world settings: Kultur Komitee Winterthur 2024 (KK24) and vTaiwan. The contributions of this paper are: (1) We introduce \textbf{Preference-based Clustering for Deliberation (PCD)}, a proposed framework that utilises voting data to create more effective deliberation groups. The process focuses on using opinion clustering to group participants of similar and dissimilar preferences together to create different deliberation dynamics. (2) We extend the \textbf{Method of Equal Shares (MES)} with a \textbf{Human-in-the-Loop} approach, allowing participants to adjust the overall budget and explore funding scenarios, balancing algorithmic decision-making with deliberation. (3) We present the \textbf{ReadTheRoom} deliberation method, which uses opinion mapping to identify divisive statements and spectrum-based voting to visualise shifts in preferences. This feedback loop fosters reflection, collaboration, and open-minded discussions. (4) We document two participatory case studies. KK24 demonstrates the impact of a \textit{Budget Assembly} for inclusive funding decisions, while vTaiwan shows how combining online and offline deliberation enhances participation and mutual learning.

\subsection{Case Studies}

\subsubsection{\textbf{Kultur Komitee: Democratising Art and Culture through Budgeting Assembly}}

Established in 2019, the \textit{Kultur Komitee} [\enquote{Culture Committee}] in Winterthur, Switzerland, empowers citizens to allocate cultural funding and shape the city’s cultural landscape. Combining elements of participatory budgeting and citizens' assemblies, this new \textit{Budget Assembly} integrates face-to-face deliberations with the tangible outcomes of participatory decision-making.

The 2023/2024 cycle (KK24) marked the fourth iteration of the \textit{Kultur Komitee} process, involving 37 randomly selected citizens. Invitations were distributed via the city’s postal service to ensure diverse participation, with 300 invitations sent and 37 accepted. The process began with a kick-off event in September 2023 where committee members developed shared goals and evaluation criteria. An open call for proposals followed, resulting in 134 submissions, which were refined into 56 shortlisted projects for deliberation. In March 2024, participants reviewed these projects individually via approval voting on an online platform, ensuring a streamlined deliberation process while capturing individual preferences. Finally, on 13 April 2024, the committee met in Winterthur to finalise funding decisions using the Human-in-the-Loop \textit{Method of Equal Shares} (MES) combined with in-person deliberations. CHF 381,500 was ultimately awarded to selected projects. This paper focuses on the later stage of the process involving online voting and the final in-person deliberation.

Out of the 37 participants involved in the process, 35 attended the final deliberative workshop (\(N = 35\)), with a mean age of 41.2 years, ranging from 20 to 82. The gender distribution consisted of 13 males and 22 females. Most participants were born in Switzerland, while 8 were from outside Switzerland. By the end of the process, 35 projects were funded, receiving grants ranging from CHF 2,500 to CHF 30,000, totalling CHF 381,500. During the process of Human-in-the-loop MES, it was decided that 50\% of the budget would be used for MES calculation, and the other half would be allocated with deliberation. Eventually, 18 projects were selected using MES, and 17 projects through deliberation. KK24 serves as a model for integrating participatory and deliberative practices in resource allocation. The authors were invited to implement the MES calculation and co-design the process with the organising committee. For more details, we refer to the KK24 factsheet \footnote{https://kulturkomitee.win/media/digital\_kk23\_24\_factsheet.pdf}.

\subsubsection{\textbf{vTaiwan: Citizen-led Deliberation Process for Public Issues}}

Launched in 2014, vTaiwan is a decentralised open consultation process combining online and offline interactions to connect citizens and government for deliberation on national issues. It serves as a model for involving diverse stakeholders in crafting digital legislation, using tools like \textit{Polis}~\cite{small2021} to facilitate large-scale conversations and achieve consensus. Since inception, vTaiwan has engaged over 200,000 participants and contributed to 26 pieces of legislation.
In December 2024, vTaiwan and the Taiwan Network Information Center co-hosted a roundtable on AI regulation, gathering input for Taiwan's proposed Basic Law on Artificial Intelligence. Using Polis and Mentimeter, the event visualised public sentiment, highlighted areas of agreement and disagreement, and tracked opinion shifts during deliberation. The corresponding author offered the ReadTheRoom Deliberation method concept through participation in vTaiwan's meetings but primarily remained as observers, analysing only anonymised open data published through vTaiwan's public repositories. This dual role as both community member and researcher is acknowledged as providing valuable insider perspective while potentially introducing bias, addressed through transparent methodology and reliance on publicly available data.

\section{Literature Review}

\subsection{Hybrid Democratic Innovation}
Hybrid Democratic Innovations (HDIs) combine concentrated deliberation with large-scale voting \cite{hendriks2024} to enhance inclusivity and policy impact. \citet{fishkin1991} pioneered this approach with deliberative polling, which \citet{mansbridge2010} describes as capturing informed public opinion through random sampling and structured deliberation. \citet{felicetti2021} expanded this concept with \enquote{democratic assemblage,} emphasising systemic integration of deliberative and participatory processes. \citet{gastil2021} demonstrated how digital platforms enhance participatory budgeting through engagement.
Practical applications include \citet{Itten2022}'s three-step climate policymaking model combining mini-publics and maxi-publics, and \citet{hendriks2024}'s analysis of Antwerp's Citizens' Budget showing how \enquote{participatory budgeting new-style} aligns deliberation with voting. For referendums, \citet{witting2023} show how deliberation addresses process deficits, while \citet{hendriks2023} propose deliberative referenda connecting citizens' assemblies with direct voting. Finally, \citet{chambers2023} theorises how voting complements deliberation by providing closure, equality, and accountability, highlighting the importance of hybrid approaches.

\subsection{Online Tools for Deliberative Democracy}
Digital tools have emerged as critical innovations in deliberative democracy, enabling scalable citizen participation in decision-making~\cite{klein_towards_2017,davies_online_2020,mikhaylovskaya_enhancing_2024}. A prominent example is Polis~\cite{small2021}, an open-source platform where participants rate statements on given topics. The tool visualises opinion spaces and computes consensus statements to foster common ground in deliberation. Used in initiatives like vTaiwan and the German \enquote{Aufstehen} movement, Polis demonstrates how digital tools can scale deliberation beyond traditional settings. Similarly, the Stanford Deliberation Platform~\cite{fishkin_deliberative_2019} uses automatic moderation for fair small-group video deliberations, as seen in the \enquote{America in One Room} project.
For discussion summarisation, tools like Wikum~\cite{zhang_wikum_2017} and Kialo~\cite{chaudoin_revolutionizing_2017} offer nested argumentation features. Decision-making platforms such as Decidim~\cite{aragon_deliberative_2017} facilitate city-level participation, while Decide Madrid and vTaiwan exemplify algorithmic empowerment for pluralistic policymaking~\cite{tseng2022}. \citet{umbelino2021}'s ProtoTeams demonstrates how team composition influences decision-making outcomes through gamified group formation.
However, scaling deliberation presents challenges, as many platforms prioritise technical solutions while overlooking cultural and linguistic diversity, social inequalities, and inclusion across different demographics~\cite{shortall_reason_2022}. Addressing these aspects is essential for supporting a broader range of participants.

\subsection{Voting and Participatory Budgeting}
Computational social choice provides tools for designing voting systems that address fairness, proportionality, and collective decision-making in participatory budgeting (PB) \cite{brandt2016}. Fairness here means equal representation of diverse preferences and proportional resource allocation \cite{fain2016}. \citet{aziz2020} survey PB models, highlighting preference elicitation, voter incentives, and welfare objectives. In multi-winner voting, \citet{brill2023} proposes robust axioms for proportional representation, while \citet{faliszewski2023} demonstrates proportional rules' utility in PB elections. \citet{peters2021} formalises proportional representation axioms in PB, proposing the \textbf{Method of Equal Shares (MES)} for fairness and efficiency. With MES, each voter receives an equal budget share usable only for projects they support. The method evaluates projects by vote count, selecting those fundable using supporters' budget shares, then dividing costs equally among those supporters. \citet{faliszewski2023} shows MES achieves fairer outcomes and improves utility. In \citet{yang2024}'s experiment, participants perceive MES as significantly fairer than conventional Greedy voting. Our paper explores how combining deliberation with digital tools for democratic decision-making can achieve proportional fairness, transparency, and practicality.

\section{Methodology}
This study employs a mixed methods approach, combining computational techniques, digital tools, and participatory frameworks to explore the integration of voting and deliberation. Our methodology is structured around three algorithmic methods: Preference-based Clustering for Deliberation, Human-in-the-loop MES, and ReadTheRoom Deliberation. As these methods have been co-designed and co-developed specifically to address the real-world issues and needs of KK24 and vTaiwan, they are also applied and evaluated in the context of these two case studies. Also, the real-world implementation of these democratic innovations required pragmatic evaluation approaches that could capture both objective outcomes and participant experiences. For each method, we developed context-specific metrics that balanced theoretical rigour with practical measurement constraints in field settings.

\subsection{Preference-based Clustering for Deliberation (PCD)}

\subsubsection{\textbf{Motivation}}

Traditional deliberative processes typically treat voting and deliberation as separate mechanisms rather than complementary parts of a unified decision-making system. This separation means valuable information from prior voting is often underutilised when forming discussion groups. The proposed Preference-based Clustering for Deliberation (PCD) framework bridges this gap by using voting data to create more effective deliberation groups.

Group deliberation often leads to middle-ground solutions that fail to address niche interests, such as local or specific community needs. This tendency for deliberation to gravitate towards consensus solutions has been noted in the literature, with scholars arguing that structured approaches are needed to ensure the representation of marginalised or niche perspectives~\cite{abdullah2016, karpowitz2014, sunstein2017}.

The theoretical foundations for this approach draw from several key concepts in deliberative democracy. \citet{fraser1990} highlights the importance of creating spaces for marginalised groups to deliberate autonomously, while \citet{mansbridge1994} emphasises the role of ``enclaves of protected discourse'' to empower these groups in broader discussions. \citet{sunstein2017} introduces the term ``\textbf{Enclave Deliberation},'' noting both its potential to address inequalities and the risks of group polarisation, which can be mitigated through careful integration into wider deliberative processes.

\subsubsection{\textbf{Process and Implementation}}
The PCD framework implements these ideas through a three-phase process: Initially, all participants cast their votes on the projects, proposals or opinions under consideration. These voting patterns are then used to inform the subsequent deliberation phases. The deliberation unfolds in two rounds: 

(1) First, participants with similar voting patterns are grouped together, creating \textbf{homogeneous deliberation} spaces where niche interests can be fully explored and articulated without being overshadowed by majority preferences. 

(2) Then, participants are reconfigured into diverse groups with different voting patterns, allowing for \textbf{heterogeneous deliberation} that builds broader consensus while ensuring minority viewpoints have already been developed. This balanced approach ensures both the depth of preference exploration and the breadth of inclusive decision-making across the deliberative process.

\begin{figure}[ht]
    \centering
    \includegraphics[width=\linewidth]{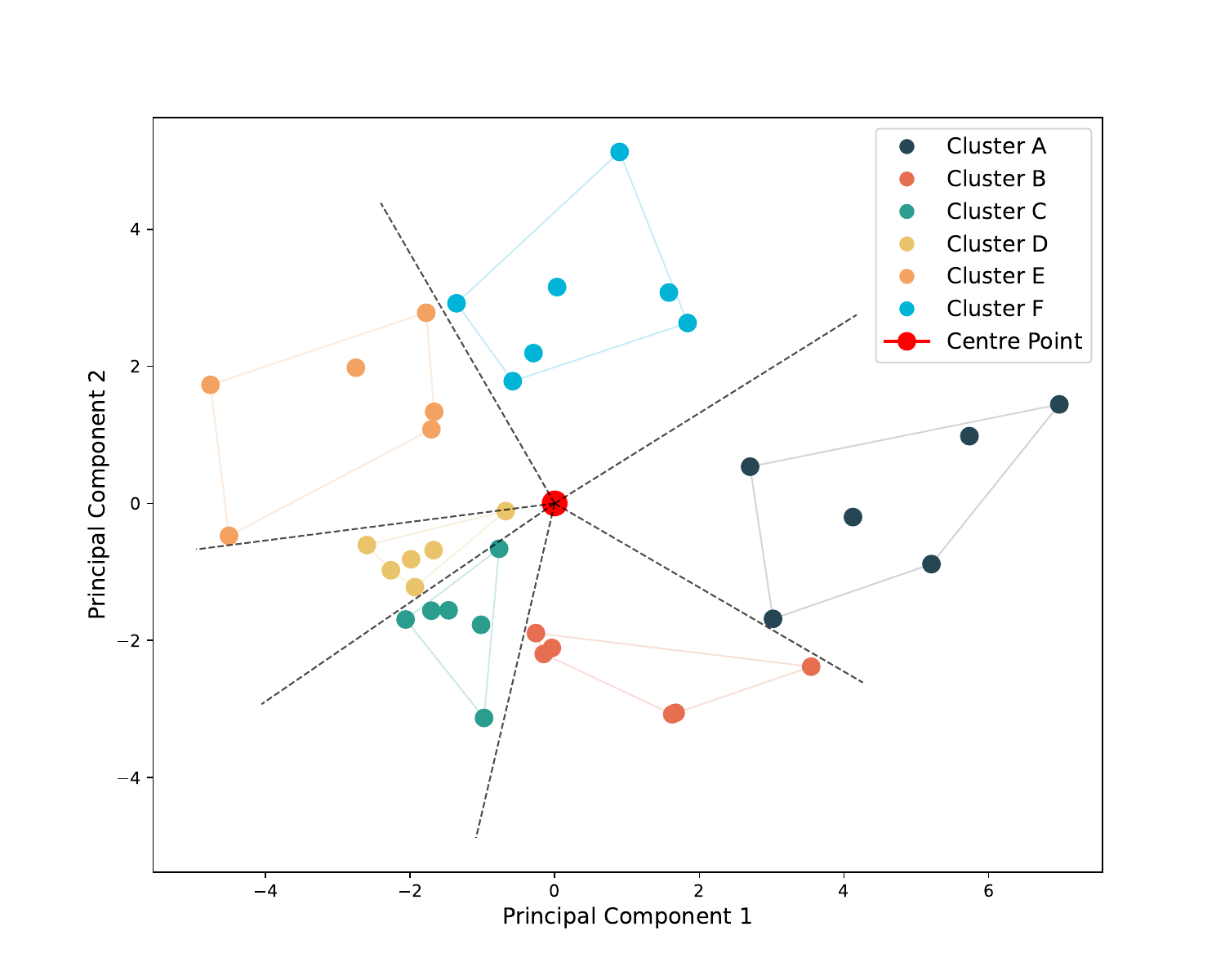}
    \caption{Radial Clustering method applied to group participants for deliberation. \normalfont\small The method projects participants into a two-dimensional opinion space using PCA and then divides them into pizza-slice-shaped sectors radiating from the center point. Each sector contains an equal number of participants, with sector angles adjusted to ensure balanced group sizes. This approach creates homogeneous deliberation groups based on voting similarity while maintaining a visually intuitive clustering representation.}
    \label{fig:voter_clusters}
\end{figure}

While the PCD framework can be implemented using various clustering algorithms, for the KK24 case study we employed a Radial Clustering method. This represents just one of many possible approaches to operationalise the framework's core principle. However, most standard clustering algorithms like k-Means~\cite{macqueen1967some} create uneven groups or exclude outliers. While balanced variants exist~\cite{franti_balanced_2014}, they can be difficult for non-technical facilitators to work with. The Radial Clustering method addresses these challenges by projecting participants into a two-dimensional opinion space and dividing it into ``pizza slice'' sectors (see Figure~\ref{fig:voter_clusters}). Each sector contains an equal number of participants, ensuring balanced group sizes.

To demonstrate this, we evaluated six clustering approaches using the KK24 pre-deliberation voting data, as detailed in Table~\ref{tab:clustering_comparison} in the appendix. The results show trade-offs between mathematical optimality and practical usability. Radial Clustering achieves a reasonable silhouette score (0.238) while maintaining perfectly balanced groups and clear visual separation between clusters. This visual simplicity offers practical advantages in real-world settings—when participants cannot attend the deliberation event, facilitators can quickly adjust the sector boundaries on printed diagrams. The broader PCD framework remains implementation agnostic, organisations can select the specific clustering algorithm that best suits their technical capabilities and deliberative context. The essential innovation lies in the PCD framework, systematically incorporating voting information to enhance deliberative processes and creating a more integrated approach to collective decision-making.

\subsubsection{\textbf{Process with Radial Clustering}}
In KK24, the organising committee determined that six groups with 6-7 participants each were needed for effective deliberation. We implemented Radial Clustering through the following steps:

(1) \textbf{Dimensionality Reduction and Mapping:} Group composition was based on participants' voting patterns from the pre-deliberation survey (approval votes for 50+ projects). Using Principal Component Analysis (PCA), we reduced this high-dimensional data to a two-dimensional space and calculated each participant's angular position relative to the mean using $\theta = \arctan \left( \frac{PC2_i - \text{mean}(PC2)}{PC1_i - \text{mean}(PC1)} \right)$, where $PC1_i$ and $PC2_i$ represent the coordinates of participant $i$. The angles $\theta$ were normalised to 0-360 degrees.

(2) \textbf{Radial Partitioning and Group Assignment:} We divided the angular space $[0, 360^\circ]$ into six sectors. Each sector was initially spanning $\frac{360}{6}$ degrees. Participants were assigned to sectors based on their angular positions, then sector boundaries were iteratively adjusted to ensure balanced group sizes of 6-7 people per group.

(3) \textbf{Two-Round Deliberation:} We conducted two deliberation rounds with different group compositions. For the first round, participants were placed in six homogeneous groups (A/B/C/D/E/F) with others sharing similar voting patterns. In the second round, they were reconfigured into six heterogeneous groups (1/2/3/4/5/6), ensuring diversity by mixing members from different homogeneous groups. In both rounds, groups discussed, selected and ranked five projects, with rankings converted to Borda scores (e.g., 1st: 5 points, 2nd: 4 points). Projects were selected based on aggregate points from all groups across both rounds until the budget was fully allocated.

\subsubsection{\textbf{Evaluation Metrics}}

To evaluate the PCD framework, we measured: \textbf{preference alignment} ($A_i = \frac{\sum_{j \in P_i} w_j}{|S_i|}$), quantifying how well group decisions reflected individual preferences. This alignment was calculated as the proportion of projects that both appeared in a participant's individual votes and received points from their deliberation group; \textbf{voting-deliberation correlation}, analysing statistical relationships between pre-deliberation votes and group decisions; \textbf{resource allocation patterns}, comparing project cost distributions between rounds using Mann-Whitney U Tests; \textbf{demographic effects}, examining alignment differences across gender and age groups with Wilcoxon signed-rank tests; and \textbf{participant experience}, using Likert-scale ratings and qualitative feedback on decision-making ease and preference representation. This approach captured both objective outcomes and subjective experiences across different group compositions.

\subsection{Human-in-the-Loop Method of Equal Shares}

\subsubsection{\textbf{Motivation}}
The Human-in-the-Loop Method of Equal Shares is a practical extension of the Method of Equal Shares algorithm, co-designed with KK24 for their needs, to enhance transparency and build algorithmic trust by incorporating real-time digital feedback. This method allows participants to control the extent to which decision-making is delegated to the voting aggregation algorithm versus reserved for deliberation. 

\subsubsection{\textbf{Method of Equal Shares (MES)\cite{peters2021}}}
The Greedy method commonly used in participatory budgeting (PB) selects projects with the highest votes but ignores proportionality, favouring majority preferences while overlooking minority groups. Research by \citet{yang_designing_2024} shows participants in PB settings are rarely cost-conscious, leading to projects winning disproportionate budget shares relative to their votes. The Method of Equal Shares (MES) \cite{peters2021} ensures fair resource allocation by assigning each voter $i \in N$ an equal starting budget $b_i \geq 0$. Here, $u_i(p)$ represents the utility voter $i$ derives from project $p$. MES operates through two key steps: \textbf{(1) Check Affordability.} A project $p$ is $q$-affordable if its cost $c_p$ can be covered by voters contributing proportionally to their utility, with each contribution capped at $q \cdot u_i(p)$. \textbf{(2) Select and Update.} The most affordable project is added to the winner set $W$, and each voter's remaining budget is updated accordingly. The algorithm begins with budget shares of $B/n$ (total budget divided by number of voters) and iteratively selects projects until no more are affordable. This approach guarantees proportionality, ensuring groups with shared preferences receive fair representation in the final project selection.

\subsubsection{\textbf{The Human-in-the-Loop extension}} 

The Human-in-the-Loop MES process begins by aggregating votes using the MES algorithm applied to a partial budget. An interactive interface enables participants to adjust a slider controlling the total budget allocation from 0 to 380,000 CHF. This feature provides real-time visualisation of the projects that would be funded under different budget scenarios using MES calculations. After exploring the outcomes, participants collaboratively decide on a specific budget allocation for the MES calculation. In this study, 190,000 CHF --- 50\% of the total budget --- was allocated to the MES calculation, resulting in 18 projects being funded. 

Following the initial selection of projects based on the MES algorithm, participants engage in a deliberation round. During this round, they discuss whether the budget of any selected projects should be adjusted or whether any projects from the MES set should be eliminated. This additional step ensures that the final set of projects aligns with participant preferences and collective priorities, providing a balance between algorithmic decision-making and human judgment.

\subsubsection{\textbf{Evaluation Metrics}}

To evaluate Human-in-the-Loop MES, we employed metrics addressing both algorithmic fairness and human agency: \textbf{project representation} measured the number of projects won per voter across methods; \textbf{budget allocation fairness} was quantified using the Gini coefficient ($G = \frac{\sum_{i=1}^{n}\sum_{j=1}^{n}|x_i - x_j|}{2n^2\bar{x}}$), with lower values indicating more equitable distribution; \textbf{perceived fairness} was assessed through participant survey responses on a 5 point Likert scale; and \textbf{human agency} was measured by examining both the percentage of algorithmic decisions modified through deliberation and participant preferences for the voting-to-deliberation ratio. 

\subsection{ReadTheRoom Deliberation}

\subsubsection{\textbf{Motivation}} Conventional deliberations often lack concrete, data-supported, and actionable outcomes for effective policymaking. Discussions can become directionless, reiterating widely accepted opinions or focusing on tangential topics, thereby undervaluing citizens' time and effort. Moreover, the process designs frequently assume static opinions, failing to capture mutual learning and opinion shifts during deliberations. The proposed \textit{ReadTheRoom} deliberation introduces a structured approach to document opinion changes, providing insights into how public sentiment evolves through discussion and evidence-based results for policy-making.

\subsubsection{\textbf{Online Phase: Wiki-Survey with Polis}} 

The \textit{Online Phase} uses \textbf{Polis}, a participatory wiki-survey platform, to enable scalable deliberation. Participants can submit statements and vote by agreeing, disagreeing, or abstaining, fostering the identification of consensus and divisive opinions. As described in \textit{The Computational Democracy Project} webpage \footnote{https://compdemocracy.org/algorithms/} and work by \citet{small2021}, Polis uses clustering algorithms to analyse responses and generate real-time reports. These reports can include a low-dimensional visualization of the collected data and several summary statistics of the debate. For example, the final report  in the AI regulation deliberation in vTaiwan\footnote{https://polis.tw/report/r3dvith8ntmwywyf4nctc} identifies 5 opinion groups, and highlights 2 divisive and 14 consensus statements.

\subsubsection{\textbf{Offline Phase: Decision Tree and Face-to-Face Deliberation}}

Building on the online phase results, our offline phase integrated digital insights into structured physical deliberations. Co-designed with vTaiwan, this approach ensured evidence-based discussions while facilitating direct learning and opinion refinement.

\textbf{Decision Tree Creation:} We used clusters and divisive statements from the Polis report to construct a visual decision tree (Figure ~\ref{fig:tree}), mapping how participants divided on key issues. This served as a discussion roadmap, making complex opinion patterns accessible to moderators and participants.

\textbf{Deliberation Process:} The structured event followed three steps: First, participants used Mentimeter for pre-discussion voting on divisive statements using a 5-point Likert scale. Second, moderators facilitated dialogue by inviting participants from diverse opinion groups to share their perspectives, encouraging mutual understanding. Finally, participants re-voted on the same statements and reflected on reasons for any changes in their positions.

In this iteration, 104 participants contributed to the online Polis survey by 17th December, while 44 attended the physical workshop on 20th December in Taipei. The groups partially overlapped, with some workshop attendees not participating in the online phase. The phases operated independently, with online results informing but not constraining the physical deliberation agenda. No demographic data was collected.

\subsubsection{\textbf{Evaluation Metrics}}

To evaluate deliberation's impact, we employed four key metrics: \textbf{percentage of voters changed}, measured as the percentage of voters who changed their stance after deliberation; \textbf{polarisation reduction}, quantified using the Bimodality Coefficient (BC = $\frac{\text{skewness}^2+1}{\text{kurtosis}+3}$), where values exceeding 0.555 indicate opinion polarisation \cite{knapp2007}; \textbf{consensus building}, calculated as $\frac{1}{1+\text{standard deviation}}$, with higher values indicating greater agreement; and \textbf{mean opinion changes}, measuring directional shifts in average responses on the 5-point Likert scale (-2 to +2). Statistical significance was assessed using paired t-tests for mean changes. 

\section{Data and Implementation}
We employed mixed methods to analyse participatory decision-making processes, combining quantitative voting data with qualitative feedback. This approach provided insights into voting patterns, algorithmic performance, and participant experiences. German and Traditional Chinese responses were translated using DeepL to facilitate analysis.

As researchers engaging with these democratic innovations, we adopted distinct roles in each case study. For KK24, we served as academic advisors, co-designing the process with the organising committee through multiple meetings from late 2023 to 2024. We provided methodological guidance, built interfaces and offered technical implementation support while ensuring decision-making autonomy remained with the organisers and participants. In the vTaiwan case, the first author participated in the public deliberation workshops as a regular attendee while documenting the process, using only data that was already publicly available online. This approach allowed for observation of deliberative dynamics firsthand while maintaining analytical distance.

All data analysed in this study are secondary, anonymised, and collected by the respective organising committees. KK24 data were gathered with signed participant consent following comprehensive briefings on data use and privacy rights. vTaiwan data contained no personal information and were collected with participant consent through Polis and Mentimeter platforms. Both datasets are publicly available for research transparency and replication\footnote{KK24 dataset: github.com/joshuay1/kk24-deliberation-voting-dataset}\footnote{vTaiwan dataset: github.com/v-taiwan/241220-AI-Regulation}.

\section{Results}

\subsection{Preference-based Clustering for Deliberation (PCD)}

\subsubsection{\textbf{Heterogeneous deliberation outcome mirror the voting outcome more closely}}

\begin{figure}[tb]
    \centering
    \includegraphics[width=\linewidth]{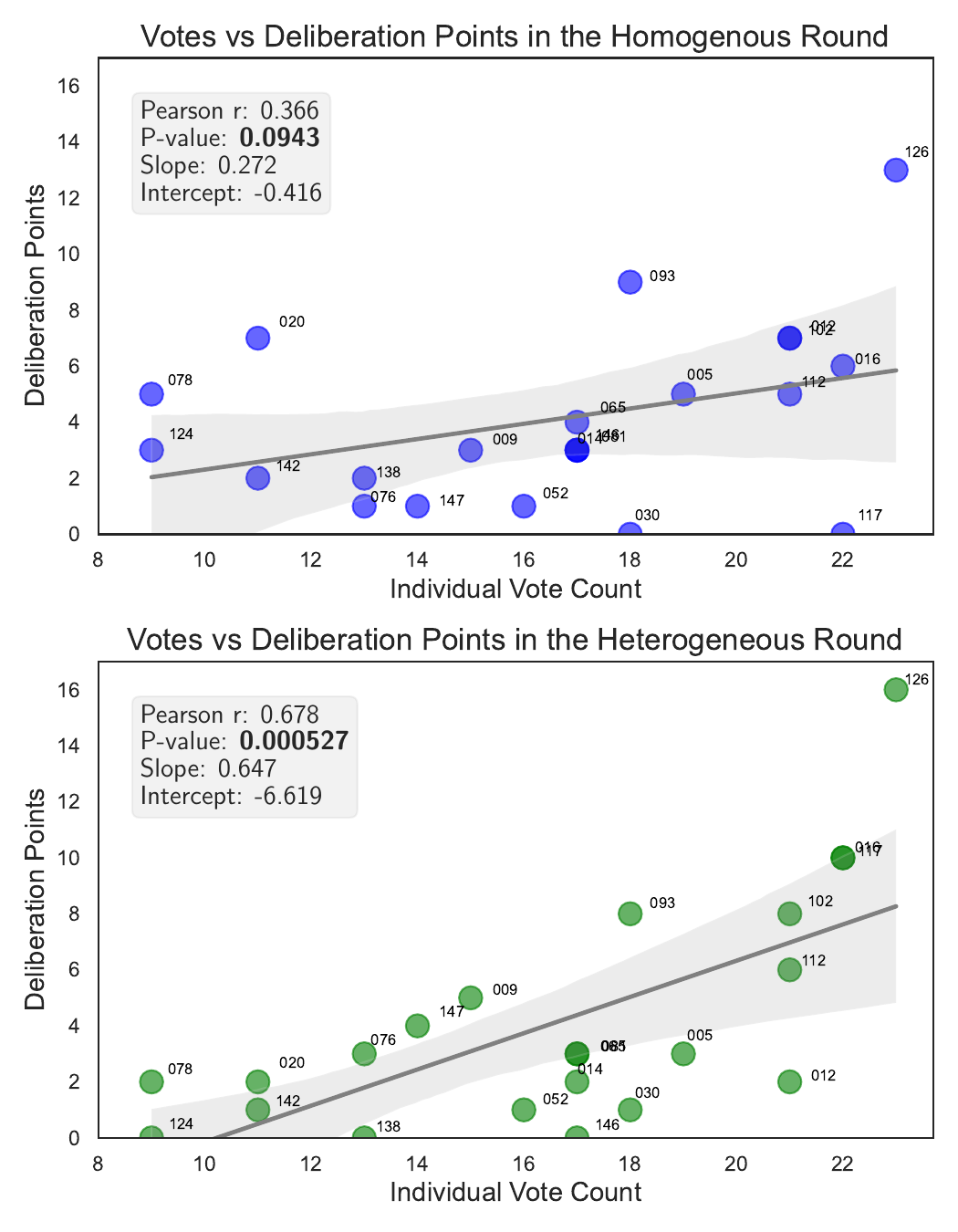}
    \caption{\textbf{Votes vs Deliberation Points in Homogeneous and Heterogeneous Rounds.} 
    \normalfont\small The relationship between the individual votes that the projects received (x-axis) and the aggregated points assigned by the groups during the physical group deliberation (y-axis). Each dot corresponds to a project. The left and right panels represent the homogeneous round and the heterogeneous round.}
    \label{fig:points_votes}
\end{figure}

The p-values in Figure~\ref{fig:points_votes} reveal a significant correlation between individual voting outcomes and group-assigned deliberation points in the heterogeneous round ($p = 0.000527$), but a non-significant correlation in the homogeneous round ($p = 0.0943$). This suggests that in heterogeneous deliberation groups, where diverse perspectives are expected to emerge, the final group decisions closely mirror individual online voting outcomes. We also observed more groups using sticky notes as votes to make collective decisions in the heterogeneous round, as shown in Figure~\ref{fig:notes}.

This pattern is further illustrated in Table~\ref{tab:final_projects} in the appendix. Among the 17 projects selected during the deliberation process, 7 overlap with those that would have been chosen if online voting alone determined the funding allocation for this portion of the budget. Compared to pure voting, heterogeneous deliberation introduces fewer unique projects. Only 3 projects were selected primarily due to heterogeneous deliberation (tagged HT), while 7 projects were selected primarily due to homogeneous deliberation (tagged HM). This suggests that homogeneous deliberation incorporates a wider range of projects that diverge from individual voting outcomes.
One possible explanation is that participants with diverse preferences in heterogeneous groups find it challenging to reach consensus, often resorting to in-group voting to resolve differences. These findings motivate the study of group dynamics in deliberative processes within systems where voting already aggregates preferences effectively.

\subsubsection{\textbf{Heterogeneous deliberation tends to fund more costly projects}}

As detailed in Table~\ref{tab:deliberation_analysis}, the hypothetical outcomes represent the project selections if the 190,000 CHF budget were allocated to projects based solely on decisions from homogeneous or heterogeneous groups. Heterogeneous deliberation funded fewer but more expensive projects than homogeneous deliberation. These findings suggest that diverse perspectives may lead to prioritising larger or higher-impact projects. 
\begin{table}[h!]
\centering
    \centering
    \begin{tabular}{l|c|c}
    \textbf{Metric} & \textbf{Homogeneous} & \textbf{Heterogeneous} \\
    \hline
    Total Budget & 186,900 & 189,900 \\
    Number of Projects & 14 & 10 \\
    \textbf{Mean Cost} & \textbf{13,350.00} & \textbf{18,990.00} \\
    Cost Std. Dev. & 8,581.71 & 12,657.05 \\
    \end{tabular}
    \caption{Comparison between hypothetical homogeneous and heterogeneous deliberation outcomes.}
    \label{tab:deliberation_analysis}
\end{table}
Although a Mann-Whitney U Test revealed that the difference in project cost distributions between the two groups is not statistically significant (U = 49.0, p = 0.22715) given the smaller data sample, the trend toward funding costlier projects under heterogeneous deliberation remains noteworthy. This observation may have important implications for resource allocation strategies in participatory processes.

\subsubsection{\textbf{More participants perceived homogeneous deliberation to be easier, with outcomes reflecting their preferences}}

As shown in Figure~\ref{fig:survey_group}, participant responses highlight notable differences in the perceived ease of decision-making and the extent to which outcomes reflected their preferences across the two rounds of deliberation. The first round, organised using Radial clustering to create homogeneous groups, was perceived as easier, with 83\% of participants agreeing that reaching a decision was easy. 65\% of the participants agreed in the second round, where Radial clustering was used to form heterogeneous groups. Similarly, decisions made by the homogeneous groups were more likely to reflect participants' preferences, with 76\% agreeing, compared to 55\% in the heterogeneous groups. These results suggest that using Radial clustering to organise groupings creates a perceivable difference for participants, with homogeneous groupings leading to smoother decision-making and outcomes more aligned with individual preferences.

\subsubsection{\textbf{Younger participants exhibit significant alignment differences between deliberation rounds}}
\begin{figure}[h]
  \includegraphics[width=\linewidth]{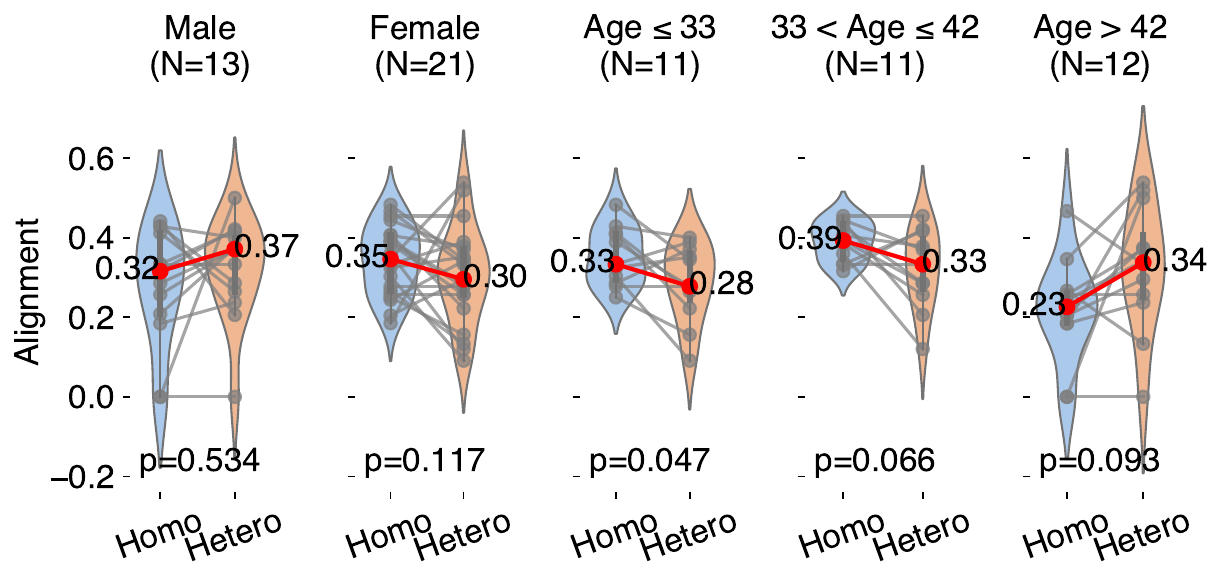}
  \caption{Alignment Changes Across Homogeneous and Heterogeneous Rounds by Gender and Age Group. \normalfont\small This figure shows the alignment of participants during the homogeneous and heterogeneous rounds. This alignment was calculated as the proportion of projects that both appeared in a participant's individual votes and received points from their deliberation group. Each subplot represents a demographic group, with the age groups created by dividing the sample into thirds. Grey lines connect individual participant alignments in the two rounds. Red dots indicate median alignments for each round, connected by a solid red line to illustrate median trends. The Wilcoxon signed-rank test p-values are displayed to assess the statistical significance of the differences between the two rounds.}
  \label{fig:socio}
\end{figure}
Figure \ref{fig:socio} compares alignment, the proportion across homogeneous and heterogeneous deliberation rounds, segmented by gender and age groups. In the homogeneous round, younger participants (Age $\leq 33$) and female participants exhibit higher median alignment (0.33 and 0.35, respectively), while older participants (Age $>$ 42) and male participants have lower alignment medians (0.23 and 0.32, respectively). In the heterogeneous round, alignment for younger participants and females decreases (to 0.28 and 0.30), whereas older participants and males experience increases (to 0.34 and 0.37, respectively). Due to the smaller sample sizes within each group, only the difference in alignment for younger participants meets the threshold for statistical significance ($p=0.047$). For the other groups, the observed differences do not reach statistical significance.

The results suggest that homogeneous deliberation may create a setting where younger and perhaps also female participants exhibit higher alignment, potentially because of fewer conflicting perspectives or reduced influence from dominant voices. However, the lower alignment observed for these groups in heterogeneous deliberations could also indicate a greater openness to new ideas or a tendency to adjust preferences in response to diverse viewpoints. These interpretations remain speculative, as the specific reasons driving these patterns are unclear and require further investigation.

\subsubsection{\textbf{The perceptions of group dynamics of deliberation were mixed}}

While most participants did not specifically comment on differences between the two rounds of deliberation, they perceived the deliberations differently. Some participants noted difficulties in ensuring equal participation within small groups. P01 observed, \textit{\enquote{When working in groups, the stronger voices sometimes had more say. I then tried to mediate, which was sometimes difficult.}} P07 echoed these challenges but appreciated the structure provided by facilitators, noting, \textit{\enquote{The main challenge to contribute my opinion that I have developed or to stand up for it / to share myself  in small groups... However, (KK24) they have made this as simple as possible with the structure.}} Others highlighted the inclusivity fostered by group dynamics. P04 remarked positively, \textit{\enquote{Through the different groups, there was always the opportunity to share one’s voice.}} Similarly, P21 appreciated the multiple avenues for participation, stating, \textit{\enquote{I find my opportunity to participate in decision-making to be appropriate overall since various opportunities for participation (voting, polls, meetings in smaller groups) were provided.}} Participants also reflected on the balance between individual preferences and collective outcomes. P16 praised the alignment between decision-making and group interests, stating, \textit{\enquote{The decision-making process was good and based on the group's own interests.}} P27 emphasised the effective integration of individual and group perspectives, describing it as a \textit{\enquote{Very good mix of individual opinions and discussion.}} While homogeneous groups may foster alignment by reducing conflicting perspectives, heterogeneous groups encourage engagement with diverse viewpoints, which may lead to mixed feelings regarding the group dynamic. Future efforts should address the influence of dominant voices while enhancing structures that support fair participation and constructive dialogue.

\begin{figure}[t]
    \centering
    \begin{minipage}[t]{0.48\linewidth}
        \centering
        \vspace{0.3cm}
        \includegraphics[width=\linewidth]{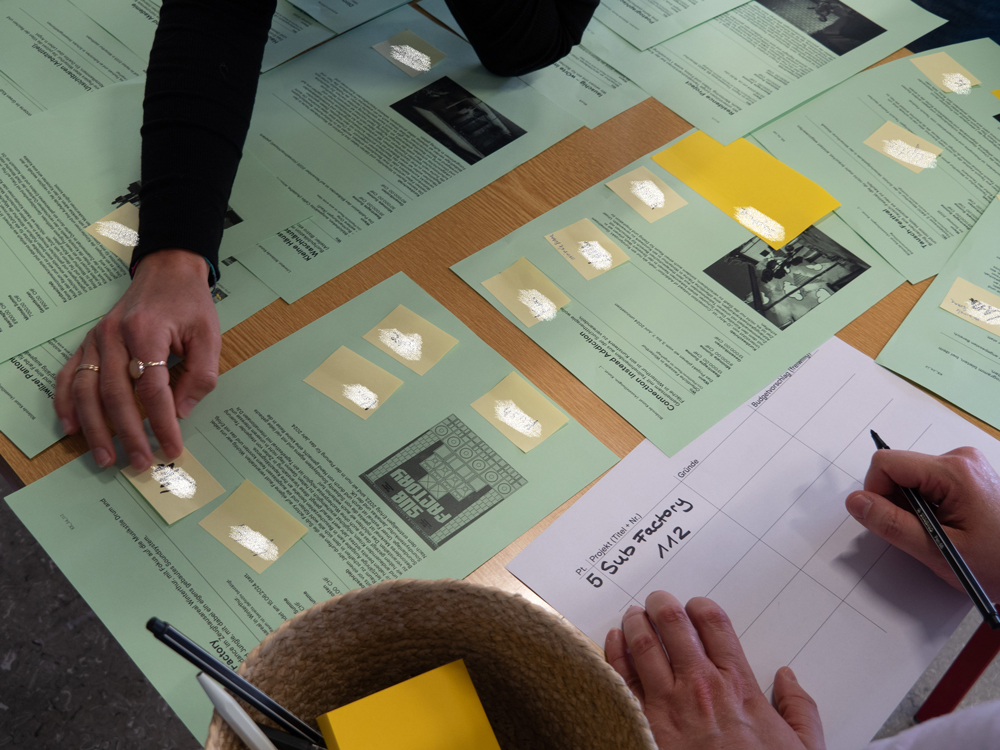}
        \caption{Voting emerged from deliberations \normalfont\small In the heterogeneous deliberation round, more participants started using the sticky notes to indicate their individual support for the project and subsequently decide what projects they should select collectively as a group.}
        \label{fig:notes}
    \end{minipage}
    \hfill
    \begin{minipage}[t]{0.48\linewidth}
        \centering
        \vspace{0.3cm}
        \includegraphics[width=\linewidth]{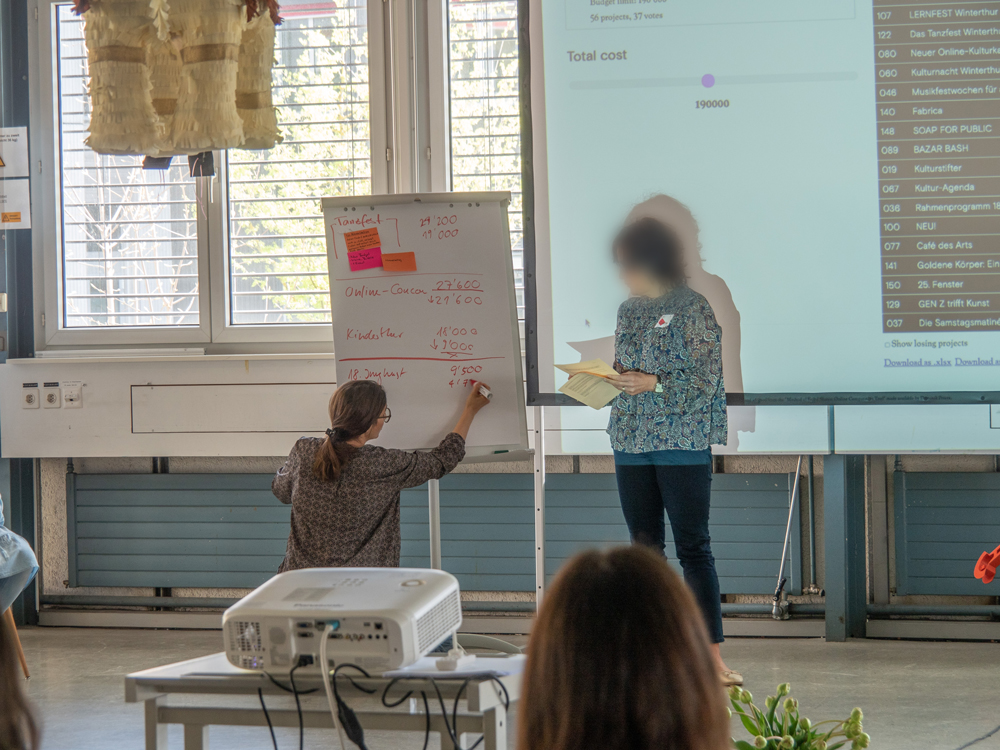}
        \caption{Adjustable budget \normalfont\small After the MES selection of projects was confirmed, participants went through the budget of each project to decide whether the proposed budget was suitable or not. They collectively voted on whether the budget should be reduced.}
        \label{fig:adjust}
    \end{minipage}
\end{figure}
\subsection{Human-in-the-loop MES}

\subsubsection{\textbf{MES distributes budget more fairly and funds more projects}}

Shown in Figure~\ref{fig:mes} in the appendix, under the same budget constraint of 190,000 CHF, the Method of Equal Shares (MES) funds more affordable projects than the conventional and commonly used Greedy Method (taking projects with the highest votes), leading to a higher mean number of projects won per voter, from 8.65 to 11.76. While the budget allocation per voter is similar in both methods, MES distributes resources more fairly, giving more to voters who receive less under the Greedy Method. This redistribution lowers the Gini coefficient from 0.17 to 0.14, reflecting a fairer outcome. The light green lines highlight these changes, showing how MES benefits voters disadvantaged by the Greedy Method while staying within the same budget.

\begin{table*}[h!]
\centering
\caption{Analysis of Deliberation Effects on Opinion Metrics}
\small
\begin{tabular}{p{4.5cm}c|ccc|ccc|ccc}
\toprule
\multirow{2}{*}{\textbf{Statement}} & \multirow{2}{*}{\textbf{Voters Changed (\%)}} & 
\multicolumn{3}{c|}{\textbf{Before Deliberation}} & \multicolumn{3}{c|}{\textbf{After Deliberation}} & \multicolumn{3}{c}{\textbf{Change}} \\
\cmidrule(lr){3-5} \cmidrule(lr){6-8} \cmidrule(lr){9-11}
& & Mean & BC & CI & Mean & BC & CI & Mean & BC & CI \\
\midrule
1. Ban AI political propaganda & 36.7 & 0.00 & 0.622 & 0.41 & -0.23 & 0.543 & 0.42 & -0.23 & -0.079 & 0.01 \\
2. AI companies should disclose & 53.1 & -0.03 & 0.561 & 0.43 & -0.34 & 0.344 & 0.50 & -0.31 & -0.217 & 0.07 \\
3. AI's industrial impact & 30.0 & -0.53 & 0.524 & 0.46 & -0.57 & 0.513 & 0.47 & -0.03 & -0.011 & 0.01 \\
4. Governments should prevent bias & 37.9 & -0.17 & 0.660 & 0.41 & 0.69 & 0.619 & 0.44 & 0.86$^*$ & -0.040 & 0.02 \\
5. AI developers must be accountable & 45.2 & -0.29 & 0.526 & 0.44 & -0.03 & 0.443 & 0.47 & 0.26 & -0.083 & 0.03 \\
\bottomrule
\multicolumn{11}{p{16cm}}{\small Note: BC = Bimodality Coefficient (values > 0.555 indicate bimodal distribution); 
CI = Consensus Index (1/(1+standard deviation)), higher values indicate more consensus; 
$^*$ indicates statistically significant change (p < 0.05).}
\end{tabular}
\label{tab:deliberation_effects}
\end{table*}

\subsubsection{\textbf{Participants value Human-in-the-loop MES for balancing algorithmic decisions and human input}}

After being shown different MES budget scenarios, participants opted to allocate 50\% of the budget using MES. The interactive interface displayed 18 selected projects, and participants were divided into randomly assigned groups to discuss potential vetoes or budget adjustments. None of the groups vetoed any projects, indicating approval of the algorithmically selected set. However, in the budget adjustment session shown in Figure~\ref{fig:adjust}, participants proposed reducing the budget for the project \enquote{Kultur-Agenda,} an online calendar for art and cultural events, from 18,000 CHF to 9,000 CHF. This additional step allowed participants to refine algorithmic decisions, balancing automation with human judgment.

Figures ~\ref{fig:survey_portion} and ~\ref{fig:survey_mes} in the appendix present participant responses to two key survey questions about MES fairness and the balance between voting and deliberation. The responses indicate that participants generally perceived MES as fair, with 62\% rating it \enquote{Very fair} and 23\% rating it \enquote{Somewhat fair.} A majority expressed support for maintaining the 50:50 ratio of individual voting to group deliberation, with an overwhelming 81\% preferring the same 50:50 proportion and only 15\% and 3\% advocating for having a higher portion of voting or deliberation, respectively.

In the open-ended feedback questions, participants expressed appreciation for the fairness and structure of MES but highlighted areas for improvement in preparation and communication. Many valued its ability to ensure all votes are counted equally, as P02 noted, \enquote{\textit{It is pragmatic and valuable to count every vote that is cast. Allows less verbal people to have equal weight given to their voice.}} Others praised its practical approach to budget management, with P09 stating, \enquote{\textit{The Method of Equal Shares provided me with a reassuring guide.}} P19 shared, \enquote{\textit{It's nice that half of the budget is distributed automatically, and the rest is then discussed and decided in detail.}} P03 commented on the balance of methods, saying, \enquote{\textit{The good thing was the mix of analogue and digital.}} P27 added, \enquote{\textit{The best thing was the mixture of individual decisions and group decisions}.} P20 suggested using more voting, stating: \enquote{\textit{More online voting; With this voting method (MES), there is no external influence.}}). While participants’ feedback did not focus much on the flexibility of Human-in-the-loop MES, since the 50:50 voting-deliberation ratio was established early on in the discussion, they frequently highlighted the value of combining voting with deliberation.

\subsection{The ReadTheRoom Deliberation Method}

\subsubsection{\textbf{Deliberation reduced polarisation and built consensus}}
Using the decision tree (Figure ~\ref{fig:tree}) in the appendix, participants deliberated on the five most divisive statements from the Polis online survey. Our analysis of before-and-after voting (Table~\ref{tab:deliberation_effects}) reveals consistent patterns across all five statements deliberated. Between 30.0\% and 53.1\% of participants changed their stance following deliberation, with the highest shift observed for Statement 2 on AI company transparency (53.1\%) and the lowest for Statement 3 on market-based solutions (30.0\%).
The Bimodality Coefficient (BC) decreased universally after deliberation, with particularly notable reductions for Statement 2 (from 0.561 to 0.344, a 38.7\% decrease) and Statement 5 on developer accountability (from 0.526 to 0.443, a 15.8\% decrease). Importantly, three statements began with BC values near or above the polarisation threshold of 0.555, but all fell below this threshold after deliberation, indicating a clear shift from divided opinion camps toward more cohesive viewpoints.
This reduction in polarisation was complemented by increases in the Consensus Index across all statements, with Statement 2 showing the largest improvement (from 0.43 to 0.50). These consistent gains in consensus suggest that deliberation effectively brought participants toward shared understanding, even on initially divisive topics.
Mean opinion values showed varied directional changes. Statement 4 on government prevention of AI bias demonstrated the most dramatic and only statistically significant shift, moving from slight disagreement (-0.17) to clear agreement (0.69). 
The limited statistical significance in our dataset (with only one change reaching p < 0.05) likely stems from our modest sample size rather than the absence of an effect. The consistency of changes across multiple metrics—decreasing BC values, increasing Consensus Indices, and substantial voter position changes—provides evidence that the ReadTheRoom method facilitates productive opinion convergence whilst preserving necessary distinctions in perspective.

\subsubsection{\textbf{Participants valued the learning experience and transparency of the process}}
Post-deliberation survey responses (Figure~\ref{fig:survey}) highlight the ReadTheRoom method's effectiveness. On a 5-point Likert scale (-2 to 2), participants gave the highest rating (1.60) to \enquote{The discussion today helped me learn new perspectives,} confirming the method created a genuine learning environment. Participants strongly agreed (1.57) that real-time voting allowed diverse views to be expressed effectively. They also valued the consensus-building process (1.13) and found the combination of voting with deliberation convincing (1.47).
Qualitative feedback reinforces these positive assessments. P05 noted, \enquote{\textit{There were many deep experiences and opinions from various fields!}}, whilst P10 shared, \enquote{\textit{I learned a lot about different positions and their concerns, and I gained a lot!}} P13 reported a clearer understanding of the issues, and several participants suggested expanding the digital tools further. P39 proposed using \enquote{\textit{more digital tools to continue the dialogue online for deeper discussions}}.
These findings demonstrate how the ReadTheRoom method transforms potentially divisive discussions into collaborative learning experiences where participants engage meaningfully with different perspectives. By making opinion shifts visible in real time, the process fosters mutual understanding whilst providing policymakers with data-supported insights into public sentiment.

\section{Discussion and Conclusion}
\subsection{Practical Impact}

This study examined three digital methods---PCD, Human-in-the-loop MES, and ReadTheRoom Deliberation---implemented in two participatory processes. While conducted on a modest scale, these methods offer innovative enhancements with broader implications.

First, the PCD framework offers a fresh approach to structuring deliberative participation. By using voting data to form discussion groups, it creates a practical bridge between individual preferences and collective decision-making. Our findings from KK24 show this method impacts both the process and outcomes in meaningful ways. Participants in homogeneous groups reported finding decisions easier to make, with outcomes better reflecting their preferences. This suggests that bringing similar voters together creates a supportive environment where niche interests can be fully expressed before broader engagement. Interestingly, our analysis showed that homogeneous groups tended to fund more numerous, smaller projects, while heterogeneous groups favoured fewer, costlier initiatives. This pattern suggests homogeneous deliberation may naturally support diverse smaller projects serving specific interests, whereas heterogeneous groups converge on larger projects with broader appeal as participants seek common ground across differing priorities. This approach aligns with Sunstein's \cite{sunstein2017} concept of enclave deliberation, while the later heterogeneous phase helps prevent potential preference polarisation. A significant strength of PCD is how it implements enclave deliberation without using demographic assumptions. As Abdullah et al. \cite{abdullah2016} emphasise, democratic enclaves should \enquote{represent marginalised perspectives or social locations rather than essentialised identities}. While we used Radial Clustering for its visual clarity and balanced groups, the PCD framework remains flexible. Different clustering methods could be selected based on specific deliberative goals and contexts. The visual simplicity of our approach had practical benefits, making the grouping logic easily understood by facilitators and participants without technical backgrounds.

Second, Human-in-the-loop MES offered a novel approach to adjusting project budgets within participatory budgeting frameworks. This approach is especially valuable for scenarios with flexible budgets, a concept rarely addressed in participatory budgeting literature due to the challenges of bridging theoretical frameworks with real-world possibilities. However, in contexts where deliberation is integral to the participatory process, Human-in-the-loop MES enables adaptable budgeting with transparent and engaging tools for monitoring and exploration. This approach could be applied to other algorithm-supported collective decisions, ensuring citizens retain control not only over outcomes but also over the extent of algorithmic involvement in the process.

Lastly, vTaiwan's {ReadTheRoom deliberation} merged the online opinion space with the physical deliberation space through a gamified spectrum, using Likert-scale questions to position participants from disagreement to agreement in two rounds: before and after the discussion. This approach ensured that participants with differing stances had a platform to voice their views while generating quantifiable data to support deliberation and provide policymakers with clear insights into citizen opinions. The interactive design encouraged participants to engage dynamically, listen, learn, and adapt their views, making deliberation both enjoyable and collaborative. By integrating real-time public input and voting data, ReadTheRoom added legitimacy to the process, enabling participants to justify their positions with evidence and giving policymakers actionable insights.

\subsection{Limitations}
Our research faced several methodological and contextual limitations. While our case studies offered valuable real-world insights, the modest sample sizes limit statistical power and generalisability. These participant numbers, though typical for deliberative mini-publics, may not capture the complexity of dynamics present in larger democratic innovations or different cultural contexts.

For the PCD framework, our findings suggest that clustering interacts with demographic factors in ways that affect deliberative outcomes. Social dynamics like confidence levels, speaking styles, and status hierarchies likely influence how different groups benefit from various deliberation structures. This makes it difficult to isolate preference-based effects from intersecting social factors. Additionally, as the KK24 outcomes represented aggregated results combining points from all groups across both deliberation rounds, it was more difficult for the participants to precisely identify the direct influence of different deliberation structures on final outcomes. Our implementation of Radial Clustering prioritised visual interpretability and practical usability over mathematical optimisation. While achieving reasonable silhouette scores, this approach potentially sacrificed precision in capturing the full dimensionality of preference spaces. The two-dimensional PCA projection, though visually intuitive, inevitably flattens the complexity of participant preferences.

With Human-in-the-loop MES, participants reviewed and modified project budgets after the MES calculation. These post-selection modifications potentially compromised MES's mathematical proportionality guarantees, creating tension between algorithmic fairness and participant agency. While such adjustments often incorporate valuable contextual knowledge, they distort the carefully calculated proportional representation that makes MES theoretically fair. Future research should develop extensions to MES that formally account for these post-selection budget modifications while preserving core fairness properties, treating such adjustments as preference updates within the algorithm itself.

As for the ReadTheRoom method, participants noted that fixed statement wording couldn't evolve alongside the discussion, potentially constraining the natural progression of ideas. The visible spectrum-based voting display might have introduced social conformity effects, with participants potentially adjusting their positions to align with perceived group norms rather than expressing independent judgments. Additionally, our evaluation captured only immediate opinion shifts during a single deliberative session, without measuring the stability of these changes over time. The observed reductions in polarisation might represent temporary accommodations rather than durable opinion changes.

Also, our evaluation approaches across all three methods prioritised practical measurement in field settings over controlled experimental design. The context-specific metrics we developed lack standardisation that would facilitate direct comparison with other deliberative innovations. Future work should develop more standardised evaluation frameworks that balance theoretical rigour with practical applicability in democratic contexts.

\subsection{Concluding Remarks}

Deliberation and voting are not opposing endpoints but complementary facets of democratic innovation \cite{weyl2024}. Our research demonstrates how computational methods can bridge deliberative depth with participatory breadth by transforming voting data from simple preference aggregation into a resource for structuring meaningful group interactions. The PCD framework, Human-in-the-loop MES, and ReadTheRoom approach collectively show how thoughtfully designed algorithms can help democratic processes overcome traditional constraints, creating decision-making systems that are simultaneously more representative, reflective, and effective as communities face increasingly complex collective challenges.

\newpage 
\clearpage

\begin{acks}

We thank the \textit{KK24 committee} and \textit{vTaiwan community}, as well as all participants, for their invaluable contributions to this research on participatory democracy. Special thanks to \textit{Noemi Scheurer} and \textit{Mia Odermatt}, the main organisers of Kultur Komitee Winterthur, for their openness to experimenting with innovative methods. We also thank \textit{Jia-Wei Cui}, the current organiser and moderator of vTaiwan, for his commitment, and \textit{Yi-Ting Lien} for co-hosting these deliberation workshops and connecting vTaiwan with TWNIC. Additionally, FB gratefully acknowledges the financial support of the Swiss National Science Foundation (SNSF) under grant ID CRSII5-205975.
\end{acks}

\bibliographystyle{ACM-Reference-Format}
\bibliography{main}


\begin{thebibliography}{41}


\ifx \showCODEN    \undefined \def \showCODEN     #1{\unskip}     \fi
\ifx \showDOI      \undefined \def \showDOI       #1{#1}\fi
\ifx \showISBNx    \undefined \def \showISBNx     #1{\unskip}     \fi
\ifx \showISBNxiii \undefined \def \showISBNxiii  #1{\unskip}     \fi
\ifx \showISSN     \undefined \def \showISSN      #1{\unskip}     \fi
\ifx \showLCCN     \undefined \def \showLCCN      #1{\unskip}     \fi
\ifx \shownote     \undefined \def \shownote      #1{#1}          \fi
\ifx \showarticletitle \undefined \def \showarticletitle #1{#1}   \fi
\ifx \showURL      \undefined \def \showURL       {\relax}        \fi
\providecommand\bibfield[2]{#2}
\providecommand\bibinfo[2]{#2}
\providecommand\natexlab[1]{#1}
\providecommand\showeprint[2][]{arXiv:#2}

\bibitem[Abdullah et~al\mbox{.}(2016)]%
        {abdullah2016}
\bibfield{author}{\bibinfo{person}{Carolyne Abdullah}, \bibinfo{person}{Christopher~F. Karpowitz}, {and} \bibinfo{person}{Chad Raphael}.} \bibinfo{year}{2016}\natexlab{}.
\newblock \showarticletitle{Affinity Groups, Enclave Deliberation, and Equity}.
\newblock \bibinfo{journal}{\emph{Journal of Public Deliberation}} \bibinfo{volume}{12}, \bibinfo{number}{2} (\bibinfo{year}{2016}), \bibinfo{pages}{Article 6}.
\newblock
\urldef\tempurl%
\url{https://delibdemjournal.org/article/id/530/}
\showURL{%
\tempurl}


\bibitem[Aragón et~al\mbox{.}(2017)]%
        {aragon_deliberative_2017}
\bibfield{author}{\bibinfo{person}{Pablo Aragón}, \bibinfo{person}{Andreas Kaltenbrunner}, \bibinfo{person}{Antonio Calleja-López}, \bibinfo{person}{Andrés Pereira}, \bibinfo{person}{Arnau Monterde}, \bibinfo{person}{Xabier~E. Barandiaran}, {and} \bibinfo{person}{Vicenç Gómez}.} \bibinfo{year}{2017}\natexlab{}.
\newblock \bibinfo{title}{Deliberative {Platform} {Design}: {The} case study of the online discussions in {Decidim} {Barcelona}}.
\newblock
\newblock


\bibitem[Aziz and Shah(2020)]%
        {aziz2020}
\bibfield{author}{\bibinfo{person}{Haris Aziz} {and} \bibinfo{person}{Nisarg Shah}.} \bibinfo{year}{2020}\natexlab{}.
\newblock \showarticletitle{Participatory budgeting: Models and approaches}.
\newblock \bibinfo{journal}{\emph{Pathways Between Social Science and Computational Social Science: Theories, Methods, and Interpretations}} (\bibinfo{year}{2020}), \bibinfo{pages}{215--236}.
\newblock


\bibitem[Brandt et~al\mbox{.}(2016)]%
        {brandt2016}
\bibfield{author}{\bibinfo{person}{Felix Brandt}, \bibinfo{person}{Vincent Conitzer}, \bibinfo{person}{Ulle Endriss}, \bibinfo{person}{J{\'e}r{\^o}me Lang}, {and} \bibinfo{person}{Ariel~D Procaccia}.} \bibinfo{year}{2016}\natexlab{}.
\newblock \bibinfo{booktitle}{\emph{Handbook of computational social choice}}.
\newblock \bibinfo{publisher}{Cambridge University Press}.
\newblock


\bibitem[Brill and Peters(2023)]%
        {brill2023}
\bibfield{author}{\bibinfo{person}{Markus Brill} {and} \bibinfo{person}{Jannik Peters}.} \bibinfo{year}{2023}\natexlab{}.
\newblock \showarticletitle{Robust and Verifiable Proportionality Axioms for Multiwinner Voting}. In \bibinfo{booktitle}{\emph{Proceedings of the 24th ACM Conference on Economics and Computation}} (London, United Kingdom) \emph{(\bibinfo{series}{EC '23})}. \bibinfo{publisher}{Association for Computing Machinery}, \bibinfo{address}{New York, NY, USA}, \bibinfo{pages}{301}.
\newblock
\showISBNx{9798400701047}
\urldef\tempurl%
\url{https://doi.org/10.1145/3580507.3597785}
\showDOI{\tempurl}


\bibitem[Bächtiger et~al\mbox{.}(2018)]%
        {bachtiger_deliberative_2018}
\bibfield{author}{\bibinfo{person}{Andre Bächtiger}, \bibinfo{person}{John~S. Dryzek}, \bibinfo{person}{Jane Mansbridge}, {and} \bibinfo{person}{Mark Warren}.} \bibinfo{year}{2018}\natexlab{}.
\newblock \showarticletitle{Deliberative {Democracy}: {An} {Introduction}}.
\newblock In \bibinfo{booktitle}{\emph{The {Oxford} {Handbook} of {Deliberative} {Democracy}}}. \bibinfo{publisher}{Oxford University Press}, \bibinfo{pages}{xxii--32}.
\newblock
\showISBNx{978-0-19-874736-9}


\bibitem[Chambers and Warren(2023)]%
        {chambers2023}
\bibfield{author}{\bibinfo{person}{Simone Chambers} {and} \bibinfo{person}{Mark~E. Warren}.} \bibinfo{year}{2023}\natexlab{}.
\newblock \showarticletitle{Why Deliberation and Voting Belong Together}.
\newblock \bibinfo{journal}{\emph{Res Publica}} (\bibinfo{year}{2023}), \bibinfo{pages}{1--19}.
\newblock
\urldef\tempurl%
\url{https://doi.org/10.1007/s11158-023-09635-x}
\showDOI{\tempurl}


\bibitem[Chaudoin et~al\mbox{.}(2017)]%
        {chaudoin_revolutionizing_2017}
\bibfield{author}{\bibinfo{person}{Stephen Chaudoin}, \bibinfo{person}{Jacob~N Shapiro}, {and} \bibinfo{person}{Dustin Tingley}.} \bibinfo{year}{2017}\natexlab{}.
\newblock \showarticletitle{Revolutionizing {Teaching} and {Research} with a {Structured} {Debate} {Platform}}.
\newblock \bibinfo{journal}{\emph{Journal of Political Science}}  \bibinfo{volume}{58} (\bibinfo{year}{2017}), \bibinfo{pages}{1064--1082}.
\newblock


\bibitem[Davies and Procter(2020)]%
        {davies_online_2020}
\bibfield{author}{\bibinfo{person}{Jonathan Davies} {and} \bibinfo{person}{Rob Procter}.} \bibinfo{year}{2020}\natexlab{}.
\newblock \showarticletitle{Online platforms of public participation: a deliberative democracy or a delusion?}. In \bibinfo{booktitle}{\emph{Proceedings of the 13th {International} {Conference} on {Theory} and {Practice} of {Electronic} {Governance}}}. \bibinfo{publisher}{ACM}, \bibinfo{address}{Athens Greece}, \bibinfo{pages}{746--753}.
\newblock
\showISBNx{978-1-4503-7674-7}


\bibitem[Fain et~al\mbox{.}(2016)]%
        {fain2016}
\bibfield{author}{\bibinfo{person}{Brandon Fain}, \bibinfo{person}{Ashish Goel}, {and} \bibinfo{person}{Kamesh Munagala}.} \bibinfo{year}{2016}\natexlab{}.
\newblock \showarticletitle{The Core of the Participatory Budgeting Problem}. In \bibinfo{booktitle}{\emph{Web and Internet Economics}}. \bibinfo{publisher}{Springer Berlin Heidelberg}, \bibinfo{pages}{384--399}.
\newblock
\urldef\tempurl%
\url{https://doi.org/10.1007/978-3-662-54110-4_27}
\showDOI{\tempurl}


\bibitem[Faliszewski et~al\mbox{.}(2023)]%
        {faliszewski2023}
\bibfield{author}{\bibinfo{person}{Piotr Faliszewski}, \bibinfo{person}{Jaros\l{}aw Flis}, \bibinfo{person}{Dominik Peters}, \bibinfo{person}{Grzegorz Pierczy\'{n}ski}, \bibinfo{person}{Piotr Skowron}, \bibinfo{person}{Dariusz Stolicki}, \bibinfo{person}{Stanis\l{}aw Szufa}, {and} \bibinfo{person}{Nimrod Talmon}.} \bibinfo{year}{2023}\natexlab{}.
\newblock \showarticletitle{Participatory budgeting: data, tools, and analysis}. In \bibinfo{booktitle}{\emph{Proceedings of the Thirty-Second International Joint Conference on Artificial Intelligence}} (Macao, P.R.China) \emph{(\bibinfo{series}{IJCAI '23})}. Article \bibinfo{articleno}{297}, \bibinfo{numpages}{8}~pages.
\newblock
\showISBNx{978-1-956792-03-4}


\bibitem[Felicetti(2021)]%
        {felicetti2021}
\bibfield{author}{\bibinfo{person}{Andrea Felicetti}.} \bibinfo{year}{2021}\natexlab{}.
\newblock \showarticletitle{Learning from democratic practices: New perspectives in institutional design}.
\newblock \bibinfo{journal}{\emph{The Journal of Politics}} \bibinfo{volume}{83}, \bibinfo{number}{4} (\bibinfo{year}{2021}), \bibinfo{pages}{1589--1601}.
\newblock


\bibitem[Fishkin et~al\mbox{.}(2019)]%
        {fishkin_deliberative_2019}
\bibfield{author}{\bibinfo{person}{James Fishkin}, \bibinfo{person}{Nikhil Garg}, \bibinfo{person}{Lodewijk Gelauff}, \bibinfo{person}{Ashish Goel}, \bibinfo{person}{Kamesh Munagala}, \bibinfo{person}{Alice Siu}, {and} \bibinfo{person}{Sravya Yandamuri}.} \bibinfo{year}{2019}\natexlab{}.
\newblock \showarticletitle{Deliberative {Democracy} with the {Online} {Deliberation} {Platform}}. In \bibinfo{booktitle}{\emph{The 7th {AAAI} {Conference} on {Human} {Computation} and {Crowdsourcing}}}.
\newblock


\bibitem[Fishkin(1991)]%
        {fishkin1991}
\bibfield{author}{\bibinfo{person}{James~S. Fishkin}.} \bibinfo{year}{1991}\natexlab{}.
\newblock \bibinfo{booktitle}{\emph{Democracy and Deliberation: New Directions for Democratic Reform}}.
\newblock \bibinfo{publisher}{Yale University Press}.
\newblock
\showISBNx{9780300051612}
\urldef\tempurl%
\url{http://www.jstor.org/stable/j.ctt1dt006v}
\showURL{%
\tempurl}


\bibitem[Fraser(1990)]%
        {fraser1990}
\bibfield{author}{\bibinfo{person}{Nancy Fraser}.} \bibinfo{year}{1990}\natexlab{}.
\newblock \showarticletitle{Rethinking the Public Sphere: A Contribution to the Critique of Actually Existing Democracy}.
\newblock \bibinfo{journal}{\emph{Social Text}} \bibinfo{number}{25/26} (\bibinfo{year}{1990}), \bibinfo{pages}{56--80}.
\newblock


\bibitem[Gastil and Broghammer(2021)]%
        {gastil2021}
\bibfield{author}{\bibinfo{person}{John Gastil} {and} \bibinfo{person}{Michael Broghammer}.} \bibinfo{year}{2021}\natexlab{}.
\newblock \showarticletitle{Linking Theories of Motivation, Game Mechanics, and Public Deliberation to Design an Online System for Participatory Budgeting}.
\newblock \bibinfo{journal}{\emph{Political Studies}} \bibinfo{volume}{69}, \bibinfo{number}{1} (\bibinfo{year}{2021}), \bibinfo{pages}{7--25}.
\newblock


\bibitem[Hendriks and Michels(2024)]%
        {hendriks2024}
\bibfield{author}{\bibinfo{person}{Frank Hendriks} {and} \bibinfo{person}{Anks Michels}.} \bibinfo{year}{2024}\natexlab{}.
\newblock \showarticletitle{Exploring the Democratic Merits of Hybrid Democratic Innovation. Combining Deliberation and Voting in Participatory Budgeting New Style}.
\newblock \bibinfo{journal}{\emph{International Journal of Public Administration}} \bibinfo{volume}{0}, \bibinfo{number}{0} (\bibinfo{year}{2024}), \bibinfo{pages}{1--11}.
\newblock
\urldef\tempurl%
\url{https://doi.org/10.1080/01900692.2024.2390492}
\showDOI{\tempurl}


\bibitem[Hendriks and Wagenaar(2023)]%
        {hendriks2023}
\bibfield{author}{\bibinfo{person}{Frank Hendriks} {and} \bibinfo{person}{Charlotte Wagenaar}.} \bibinfo{year}{2023}\natexlab{}.
\newblock \showarticletitle{The Deliberative Referendum: An Idea Whose Time has Come?}
\newblock \bibinfo{journal}{\emph{Administration \& Society}} \bibinfo{volume}{55}, \bibinfo{number}{3} (\bibinfo{year}{2023}), \bibinfo{pages}{569--590}.
\newblock
\urldef\tempurl%
\url{https://doi.org/10.1177/00953997221140898}
\showDOI{\tempurl}


\bibitem[Itten and Mouter(2022)]%
        {Itten2022}
\bibfield{author}{\bibinfo{person}{Anatol Itten} {and} \bibinfo{person}{Niek Mouter}.} \bibinfo{year}{2022}\natexlab{}.
\newblock \showarticletitle{When Digital Mass Participation Meets Citizen Deliberation: Combining Mini- and Maxi-Publics in Climate Policy-Making}.
\newblock \bibinfo{journal}{\emph{Sustainability}} \bibinfo{volume}{14}, \bibinfo{number}{8} (\bibinfo{year}{2022}).
\newblock
\showISSN{2071-1050}
\urldef\tempurl%
\url{https://doi.org/10.3390/su14084656}
\showDOI{\tempurl}


\bibitem[Karpowitz and Raphael(2014)]%
        {karpowitz2014}
\bibfield{author}{\bibinfo{person}{Christopher~F. Karpowitz} {and} \bibinfo{person}{Chad Raphael}.} \bibinfo{year}{2014}\natexlab{}.
\newblock \bibinfo{booktitle}{\emph{Deliberation, Democracy, and Civic Forums: Improving Equality and Publicity}}.
\newblock \bibinfo{publisher}{Cambridge University Press}, \bibinfo{address}{New York, NY}.
\newblock
\urldef\tempurl%
\url{https://www.cambridge.org/core/books/deliberation-democracy-and-civic-forums/53A4BB5D4E3BA1042BD0CB361F04D082}
\showURL{%
\tempurl}


\bibitem[Klein(2017)]%
        {klein_towards_2017}
\bibfield{author}{\bibinfo{person}{Mark Klein}.} \bibinfo{year}{2017}\natexlab{}.
\newblock \showarticletitle{Towards {Crowd}-{Scale} {Deliberation}}.
\newblock \bibinfo{journal}{\emph{SSRN Electronic Journal}} (\bibinfo{year}{2017}).
\newblock
\showISSN{1556-5068}


\bibitem[Knapp(2007)]%
        {knapp2007}
\bibfield{author}{\bibinfo{person}{Thomas~R Knapp}.} \bibinfo{year}{2007}\natexlab{}.
\newblock \showarticletitle{Bimodality revisited}.
\newblock \bibinfo{journal}{\emph{Journal of Modern Applied Statistical Methods}} \bibinfo{volume}{6}, \bibinfo{number}{1} (\bibinfo{year}{2007}), \bibinfo{pages}{3}.
\newblock


\bibitem[List(2018)]%
        {bachtiger_democratic_2018}
\bibfield{author}{\bibinfo{person}{Christian List}.} \bibinfo{year}{2018}\natexlab{}.
\newblock \showarticletitle{Democratic {Deliberation} and {Social} {Choice}: {A} {Review}}.
\newblock In \bibinfo{booktitle}{\emph{The {Oxford} {Handbook} of {Deliberative} {Democracy}}}. \bibinfo{publisher}{Oxford University Press}, \bibinfo{pages}{462--489}.
\newblock
\showISBNx{978-0-19-874736-9}


\bibitem[Mackie(2018)]%
        {bachtiger_deliberation_2018-3}
\bibfield{author}{\bibinfo{person}{Gerry Mackie}.} \bibinfo{year}{2018}\natexlab{}.
\newblock \showarticletitle{Deliberation and {Voting} {Entwined}}.
\newblock In \bibinfo{booktitle}{\emph{The {Oxford} {Handbook} of {Deliberative} {Democracy}}}. \bibinfo{publisher}{Oxford University Press}, \bibinfo{pages}{217--236}.
\newblock
\showISBNx{978-0-19-874736-9}


\bibitem[MacQueen et~al\mbox{.}(1967)]%
        {macqueen1967some}
\bibfield{author}{\bibinfo{person}{James MacQueen} {et~al\mbox{.}}} \bibinfo{year}{1967}\natexlab{}.
\newblock \showarticletitle{Some methods for classification and analysis of multivariate observations}. In \bibinfo{booktitle}{\emph{Proceedings of the fifth Berkeley symposium on mathematical statistics and probability}}, Vol.~\bibinfo{volume}{1}. Oakland, CA, USA, \bibinfo{pages}{281--297}.
\newblock


\bibitem[Malinen and Fränti(2014)]%
        {franti_balanced_2014}
\bibfield{author}{\bibinfo{person}{Mikko~I. Malinen} {and} \bibinfo{person}{Pasi Fränti}.} \bibinfo{year}{2014}\natexlab{}.
\newblock \showarticletitle{Balanced {K}-{Means} for {Clustering}}.
\newblock In \bibinfo{booktitle}{\emph{Structural, {Syntactic}, and {Statistical} {Pattern} {Recognition}}}. Vol.~\bibinfo{volume}{8621}. \bibinfo{publisher}{Springer Berlin Heidelberg}, \bibinfo{address}{Berlin, Heidelberg}, \bibinfo{pages}{32--41}.
\newblock
\showISBNx{978-3-662-44414-6 978-3-662-44415-3}


\bibitem[Mansbridge(1994)]%
        {mansbridge1994}
\bibfield{author}{\bibinfo{person}{Jane Mansbridge}.} \bibinfo{year}{1994}\natexlab{}.
\newblock \showarticletitle{Using Power/Fighting Power}.
\newblock \bibinfo{journal}{\emph{Constellations}} \bibinfo{volume}{1}, \bibinfo{number}{1} (\bibinfo{year}{1994}), \bibinfo{pages}{53--73}.
\newblock
\urldef\tempurl%
\url{https://doi.org/10.1111/j.1467-8675.1994.tb00004.x}
\showDOI{\tempurl}


\bibitem[Mansbridge(2010)]%
        {mansbridge2010}
\bibfield{author}{\bibinfo{person}{Jane Mansbridge}.} \bibinfo{year}{2010}\natexlab{}.
\newblock \showarticletitle{Deliberative Polling as the Gold Standard}.
\newblock \bibinfo{journal}{\emph{The Good Society}} \bibinfo{volume}{19}, \bibinfo{number}{1} (\bibinfo{year}{2010}), \bibinfo{pages}{55--62}.
\newblock
\urldef\tempurl%
\url{https://doi.org/10.5325/goodsociety.19.1.0055}
\showDOI{\tempurl}


\bibitem[Michels(2011)]%
        {michels2011}
\bibfield{author}{\bibinfo{person}{Ank Michels}.} \bibinfo{year}{2011}\natexlab{}.
\newblock \showarticletitle{Innovations in democratic governance: how does citizen participation contribute to a better democracy?}
\newblock \bibinfo{journal}{\emph{International Review of Administrative Sciences}} \bibinfo{volume}{77}, \bibinfo{number}{2} (\bibinfo{year}{2011}), \bibinfo{pages}{275--293}.
\newblock


\bibitem[Mikhaylovskaya(2024)]%
        {mikhaylovskaya_enhancing_2024}
\bibfield{author}{\bibinfo{person}{Anna Mikhaylovskaya}.} \bibinfo{year}{2024}\natexlab{}.
\newblock \showarticletitle{Enhancing {Deliberation} with {Digital} {Democratic} {Innovations}}.
\newblock \bibinfo{journal}{\emph{Philosophy \& Technology}} \bibinfo{volume}{37}, \bibinfo{number}{1} (\bibinfo{date}{March} \bibinfo{year}{2024}), \bibinfo{pages}{3}.
\newblock
\showISSN{2210-5433, 2210-5441}


\bibitem[Peters et~al\mbox{.}(2021)]%
        {peters2021}
\bibfield{author}{\bibinfo{person}{Dominik Peters}, \bibinfo{person}{Grzegorz Pierczynski}, {and} \bibinfo{person}{Piotr Skowron}.} \bibinfo{year}{2021}\natexlab{}.
\newblock \showarticletitle{{Proportional Participatory Budgeting with Additive Utilities}}. In \bibinfo{booktitle}{\emph{NeurIPS}}.
\newblock
\urldef\tempurl%
\url{https://proceedings.neurips.cc/paper/2021/file/69f8ea31de0c00502b2ae571fbab1f95-Paper.pdf}
\showURL{%
\tempurl}


\bibitem[Shortall et~al\mbox{.}(2022)]%
        {shortall_reason_2022}
\bibfield{author}{\bibinfo{person}{Ruth Shortall}, \bibinfo{person}{Anatol Itten}, \bibinfo{person}{Michiel Van~Der Meer}, \bibinfo{person}{Pradeep Murukannaiah}, {and} \bibinfo{person}{Catholijn Jonker}.} \bibinfo{year}{2022}\natexlab{}.
\newblock \showarticletitle{Reason against the machine? {Future} directions for mass online deliberation}.
\newblock \bibinfo{journal}{\emph{Frontiers in Political Science}}  \bibinfo{volume}{4} (\bibinfo{date}{Oct.} \bibinfo{year}{2022}), \bibinfo{pages}{946589}.
\newblock
\showISSN{2673-3145}


\bibitem[Small et~al\mbox{.}(2021)]%
        {small2021}
\bibfield{author}{\bibinfo{person}{Christopher Small}, \bibinfo{person}{Michael Bjorkegren}, \bibinfo{person}{Timo Erkkil{\"a}}, \bibinfo{person}{Lynette Shaw}, {and} \bibinfo{person}{Colin Megill}.} \bibinfo{year}{2021}\natexlab{}.
\newblock \showarticletitle{Polis: Scaling deliberation by mapping high dimensional opinion spaces}.
\newblock \bibinfo{journal}{\emph{Recerca: revista de pensament i an{\`a}lisi}} \bibinfo{volume}{26}, \bibinfo{number}{2} (\bibinfo{year}{2021}).
\newblock


\bibitem[Sunstein(2017)]%
        {sunstein2017}
\bibfield{author}{\bibinfo{person}{Cass~R Sunstein}.} \bibinfo{year}{2017}\natexlab{}.
\newblock \showarticletitle{Deliberative trouble? Why groups go to extremes}.
\newblock In \bibinfo{booktitle}{\emph{Multi-party dispute resolution, democracy and decision-making}}. \bibinfo{publisher}{Routledge}, \bibinfo{pages}{65--95}.
\newblock


\bibitem[Tseng(2022)]%
        {tseng2022}
\bibfield{author}{\bibinfo{person}{Yu-Shan Tseng}.} \bibinfo{year}{2022}\natexlab{}.
\newblock \showarticletitle{Algorithmic empowerment: A comparative ethnography of two open-source algorithmic platforms--Decide Madrid and vTaiwan}.
\newblock \bibinfo{journal}{\emph{Big Data \& Society}} \bibinfo{volume}{9}, \bibinfo{number}{2} (\bibinfo{year}{2022}), \bibinfo{pages}{20539517221123505}.
\newblock


\bibitem[Umbelino et~al\mbox{.}(2021)]%
        {umbelino2021}
\bibfield{author}{\bibinfo{person}{Gustavo Umbelino}, \bibinfo{person}{Matin Yarmand}, \bibinfo{person}{Samuel Blake}, \bibinfo{person}{Vivian Ta}, \bibinfo{person}{Amy Luo}, {and} \bibinfo{person}{Steven~P. Dow}.} \bibinfo{year}{2021}\natexlab{}.
\newblock \showarticletitle{ProtoTeams: Supporting Team Dating in Co-Located Settings}.
\newblock \bibinfo{journal}{\emph{Proceedings of the ACM on Human-Computer Interaction}} \bibinfo{volume}{4}, \bibinfo{number}{CSCW3} (\bibinfo{year}{2021}), \bibinfo{pages}{273:1--273:27}.
\newblock
\urldef\tempurl%
\url{https://doi.org/10.1145/3434182}
\showDOI{\tempurl}


\bibitem[Weyl et~al\mbox{.}(2024)]%
        {weyl2024}
\bibfield{author}{\bibinfo{person}{E.G. Weyl}, \bibinfo{person}{A. Tang}, {and} \bibinfo{person}{Community}.} \bibinfo{year}{2024}\natexlab{}.
\newblock \bibinfo{booktitle}{\emph{Plurality: The Future of Collaborative Technology and Democracy}}.
\newblock \bibinfo{publisher}{Radicalxchange}.
\newblock
\showISBNx{9798869327222}
\urldef\tempurl%
\url{https://books.google.ch/books?id=-ye20AEACAAJ}
\showURL{%
\tempurl}


\bibitem[Witting et~al\mbox{.}(2025)]%
        {witting2023}
\bibfield{author}{\bibinfo{person}{Irene Witting}, \bibinfo{person}{Charlotte Wagenaar}, {and} \bibinfo{person}{Frank Hendriks}.} \bibinfo{year}{2025}\natexlab{}.
\newblock \showarticletitle{Improving referendums with deliberative democracy: A systematic literature review}.
\newblock \bibinfo{journal}{\emph{International Political Science Review}} \bibinfo{volume}{46}, \bibinfo{number}{1} (\bibinfo{year}{2025}), \bibinfo{pages}{40--56}.
\newblock


\bibitem[Yang et~al\mbox{.}(2024a)]%
        {yang_designing_2024}
\bibfield{author}{\bibinfo{person}{Joshua~C. Yang}, \bibinfo{person}{Carina~I. Hausladen}, \bibinfo{person}{Dominik Peters}, \bibinfo{person}{Evangelos Pournaras}, \bibinfo{person}{Regula~Hänggli Fricker}, {and} \bibinfo{person}{Dirk Helbing}.} \bibinfo{year}{2024}\natexlab{a}.
\newblock \bibinfo{title}{Designing {Digital} {Voting} {Systems} for {Citizens}: {Achieving} {Fairness} and {Legitimacy} in {Participatory} {Budgeting}}.
\newblock
\newblock


\bibitem[Yang et~al\mbox{.}(2024b)]%
        {yang2024}
\bibfield{author}{\bibinfo{person}{Joshua~C. Yang}, \bibinfo{person}{Carina~I. Hausladen}, \bibinfo{person}{Dominik Peters}, \bibinfo{person}{Evangelos Pournaras}, \bibinfo{person}{Regula Hnggli~Fricker}, {and} \bibinfo{person}{Dirk Helbing}.} \bibinfo{year}{2024}\natexlab{b}.
\newblock \showarticletitle{Designing Digital Voting Systems for Citizens: Achieving Fairness and Legitimacy in Participatory Budgeting}.
\newblock \bibinfo{journal}{\emph{Digit. Gov.: Res. Pract.}} \bibinfo{volume}{5}, \bibinfo{number}{3}, Article \bibinfo{articleno}{26} (\bibinfo{date}{Sept.} \bibinfo{year}{2024}), \bibinfo{numpages}{30}~pages.
\newblock
\urldef\tempurl%
\url{https://doi.org/10.1145/3665332}
\showDOI{\tempurl}


\bibitem[Zhang et~al\mbox{.}(2017)]%
        {zhang_wikum_2017}
\bibfield{author}{\bibinfo{person}{Amy~X. Zhang}, \bibinfo{person}{Lea Verou}, {and} \bibinfo{person}{David Karger}.} \bibinfo{year}{2017}\natexlab{}.
\newblock \showarticletitle{Wikum: {Bridging} {Discussion} {Forums} and {Wikis} {Using} {Recursive} {Summarization}}. In \bibinfo{booktitle}{\emph{Proceedings of the 2017 {ACM} {Conference} on {Computer} {Supported} {Cooperative} {Work} and {Social} {Computing}}}. \bibinfo{publisher}{ACM}, \bibinfo{address}{Portland Oregon USA}, \bibinfo{pages}{2082--2096}.
\newblock
\showISBNx{978-1-4503-4335-0}


\end{thebibliography}

\appendix
\onecolumn
\section{Preference-based Clustering for Deliberation}

\begin{table}[H]
\centering
\begin{tabular}{l|c|c|c}
\hline
\textbf{Algorithm} & \textbf{SS} & \textbf{BG} & \textbf{OC} \\
\hline
Radial Clustering & 0.238 & Yes & No \\
Low-Dim Balanced K-Means & 0.220 & Yes & No \\
High-Dim Balanced K-Means & -0.051 & Yes & Yes \\
One-Dim PCA Clustering & 0.030 & Yes & No \\
Low-Dim K-Means & 0.429 & No & No \\
High-Dim K-Means & -0.175 & No & Yes \\
\hline
\\
\end{tabular}
\caption{Comparison of clustering algorithms for deliberation group formation. \small\normalfont The algorithms include our proposed Radial Clustering, standard K-Means variants in both low and high dimensions, balanced K-Means variants ensuring equal group sizes, and a one-dimensional PCA-based approach. The Silhouette Score (SS) measures cluster quality (higher is better), with calculations performed in low-dimensional space for visualization clarity. The second column (BG) indicates whether the algorithm produces equal-sized groups. The third column (OC) indicates whether the resulting clusters have overlapping clusters (i.e., in contrast to clear visual separation in low-dimensional space, which is crucial for non-technical facilitators to interpret and modify groups as needed).}
\label{tab:clustering_comparison}
\end{table}

\begin{figure}[ht]
    \vspace{0pt}
    \centering
    \includegraphics[width=0.5\linewidth]{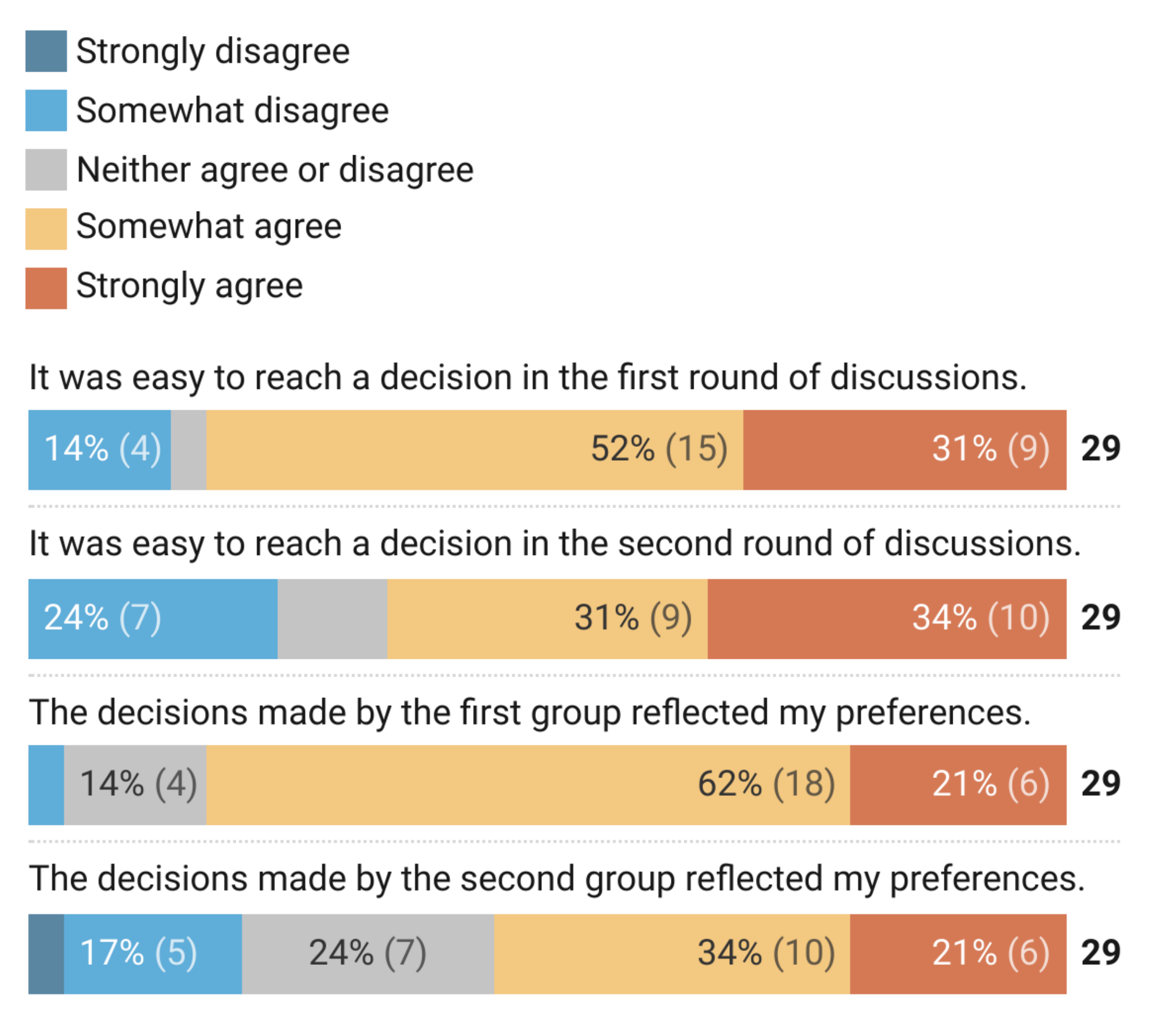}
    \caption{Participant responses on the ease of decision-making and alignment with preferences during the first (homogeneous) and second (heterogeneous) rounds of deliberation.}
    \label{fig:survey_group}
\end{figure}

\definecolor{LightBlue}{RGB}{240, 250, 240} 
\definecolor{LightRed}{RGB}{220, 240, 250}  
\definecolor{LightGrey}{RGB}{245, 245, 245} 

\onecolumn
\subsection{KK24 Outcome Table}

\begin{longtable}{>{\raggedright\arraybackslash}p{1cm} p{8cm} r r r c c}
\caption{Comparative Analysis of Projects Selected Through Deliberation vs. Standard Voting
\small\normalfont This table examines 17 cultural projects that gained funding through the deliberative process, showing how different decision-making methods affect project selection outcomes. In KK24, half the budget was allocated via pre-deliberation online voting, with the remaining half determined by deliberation points. The tick marks in the 'Vote' column indicate the 7 projects that would have received funding even without deliberation—revealing that 10 projects were funded solely because of the deliberative process. Projects are categorised based on which deliberation method awarded more points: \textbf{HM} indicates projects receiving more points from homogeneous group deliberation than from heterogeneous deliberation (7 projects); \textbf{HT} shows projects favoured by heterogeneous deliberation (3 projects); and \textbf{HM/HT} represents projects with equal points from both methods (1 project). This distribution suggests homogeneous deliberation introduces greater diversity in selected projects compared to both standard voting and heterogeneous deliberation approaches.}

\label{tab:final_projects} \\
\toprule
\textbf{ID} & \textbf{Project Name} & \textbf{Cost (CHF)} & \textbf{HM Pts} & \textbf{HT Pts} & \textbf{Tag} & \textbf{Vote} \\
\midrule
\endfirsthead

\toprule
\textbf{ID} & \textbf{Project Name} & \textbf{Cost (CHF)} & \textbf{HM Pts} & \textbf{HT Pts} & \textbf{Tag} & \textbf{Vote} \\
\midrule
\endhead

\bottomrule
\endfoot

5 & Wir, hier. Briefe an Winterthur & 24,500 & 5 & 3 & HM &  \\
14 & Haltestelle 21: Pedibus de la Culture – De Pedibus vo Winti & 20,000 & 3 & 2 & HM &  \\
20 & Cashflow & 25,000 & 7 & 2 & HM &  \\
65 & Clown-Theater Schanz \& Ganz & 20,000 & 4 & 3 & HM &  \\
78 & Aufführung des Oratoriums \enquote{Die Schöpfung} von Joseph Haydn & 5,000 & 5 & 2 & HM &  \\
93 & Gesehen & 13,000 & 9 & 8 & HM &  \\
142 & \enquote{copy paste} & 4,500 & 2 & 1 & HM &  \\

9 & F --- E Filmpreis & 28,000 & 3 & 5 & HT &  \\
76 & Peter und der Wolf – Konzertsaison 24/25 & 40,000 & 1 & 3 & HT &  \\
147 & Konzertreihe Salle Bolivar & 10,000 & 1 & 4 & HT &  \\
16 & Dokfilm: Die Unsichtbaren (Arbeitstitel). & 20,000 & 6 & 10 & HT & \checkmark \\
102 & Afro-Classics im Rahmen des Afro-Pfingsten Festivals 2025 & 8,000 & 7 & 8 & HT & \checkmark \\
117 & lauschig – wOrte im Freien 2024 & 40,000 & 0 & 10 & HT & \checkmark \\
126 & Connection instead Addiction & 13,000 & 13 & 16 & HT & \checkmark \\

81 & AlbTraumWelt & 9,900 & 3 & 3 & HM/HT & \checkmark \\

12 & Bambolini! & 5,000 & 7 & 2 & HM & \checkmark \\
112 & Sub Factory & 8,000 & 5 & 6 & HT & \checkmark \\
\end{longtable}

\section{Human-in-the-loop MES}

\begin{figure}[ht]
    \centering
    \includegraphics[width=0.5\linewidth]{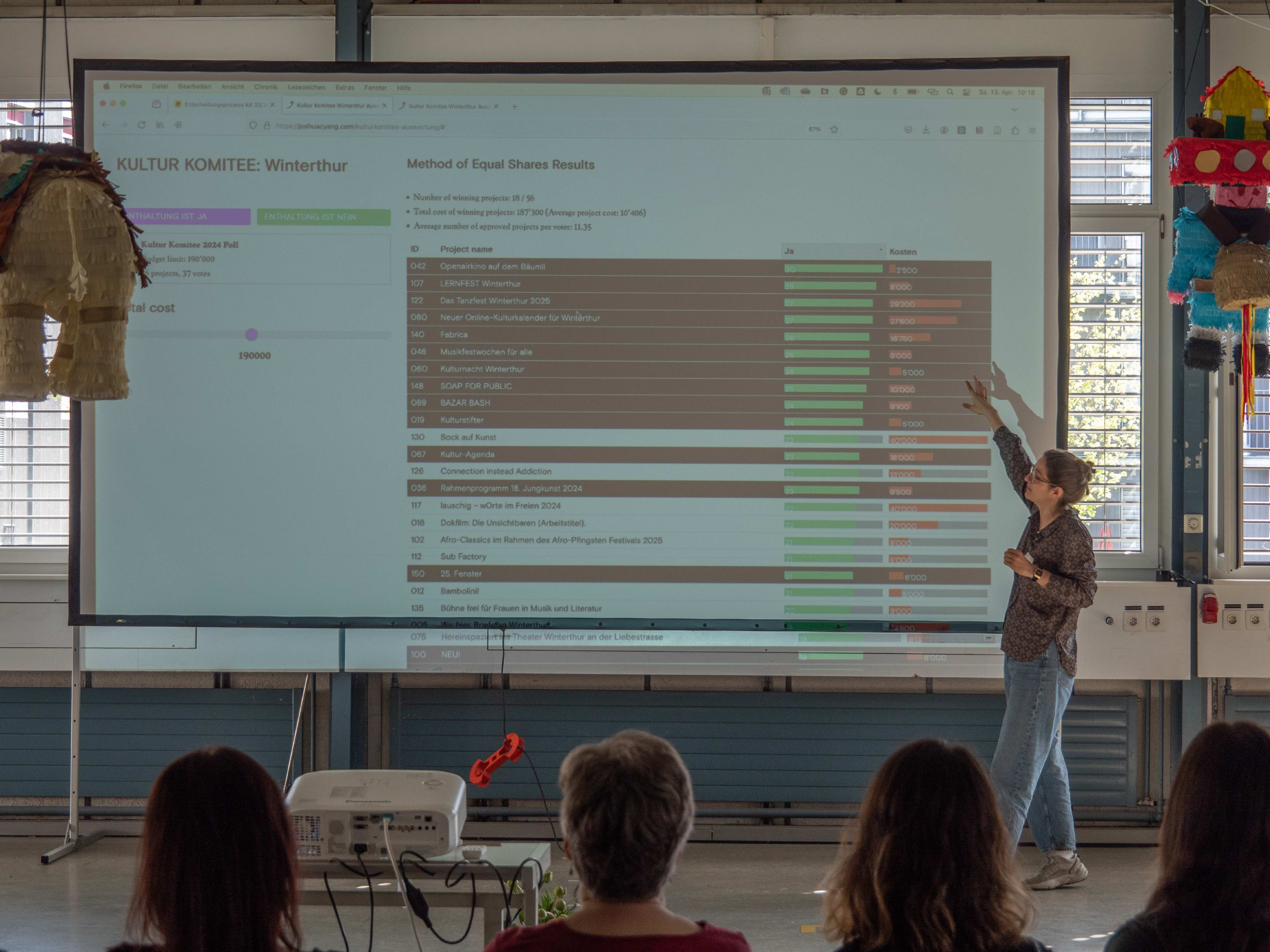}
    \caption{Real-time Human-in-the-loop MES results interface used during the KK24 workshop. \small\normalfont The total budget could be adjusted to see what projects can be funded. The green bar represents the total number of votes, and the orange bar shows the project cost. Funded projects are highlighted in black.}
    \label{fig:MES_results}
\end{figure}

\begin{figure*}[tb]
    \centering
    \includegraphics[width=\linewidth]{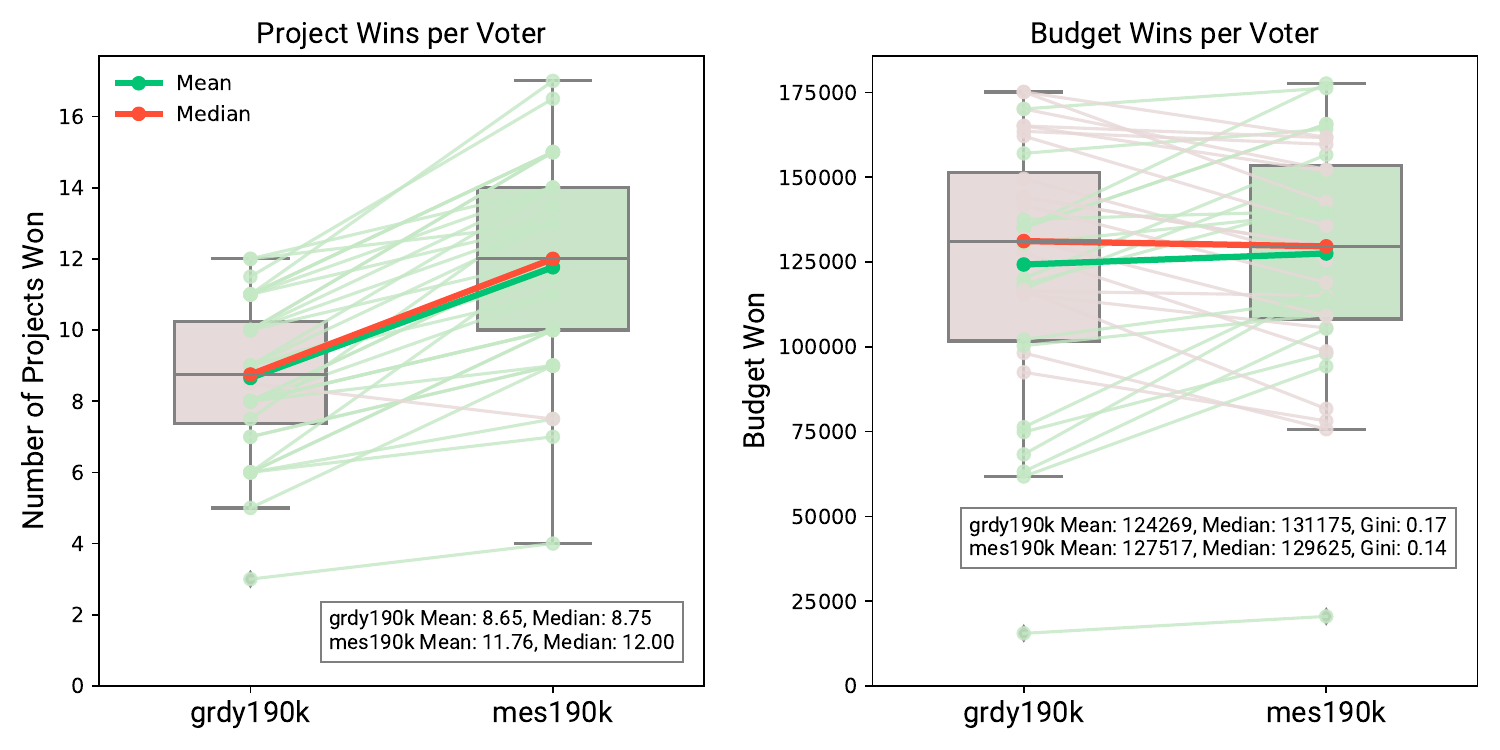}
    \caption{
        Outcome comparison of MES (\textit{mes190k}), as implemented in KK24, and a hypothetical Greedy Method baseline (\textit{grdy190k}) under a budget constraint of 190,000 CHF. \small\normalfont The left panel shows the number of projects won per voter, and the right panel shows the budget allocation per voter. Individual voter trajectories are represented by connecting lines, where light green lines indicate an increase in the outcome under MES compared to the Greedy Method, and grey lines indicate no increase. Boxplots summarise the distributions, showing the median (central line), interquartile range (box), and whiskers extending to 1.5 times the IQR or the data extremes. Mean (green) and median (red) markers are overlaid, and the Gini coefficient in the right panel quantifies the fairness of budget allocations.
    }
    \label{fig:mes}
\end{figure*}

\begin{figure}[H]
    \centering
    \includegraphics[width=0.9\linewidth]{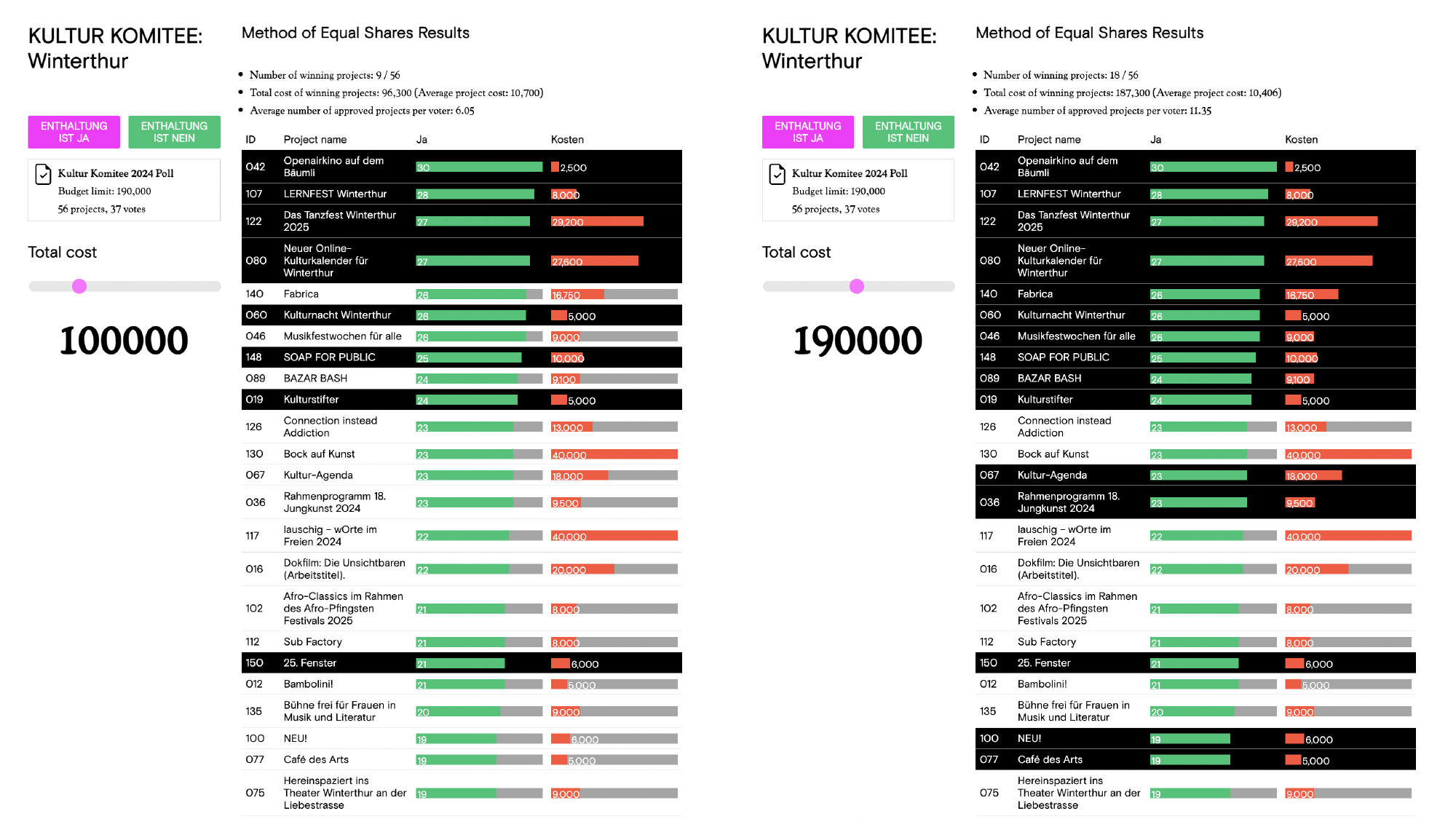}
    \caption{Screenshots of the Human-in-the-loop MES interface used in the KK24 workshop for participants to clearly understand the implications and the final selections of projects with MES calculation and ajustable budget}
    \label{fig:ui}
\end{figure}

\begin{figure}[H]

\centering
\includegraphics[width=0.5\linewidth]{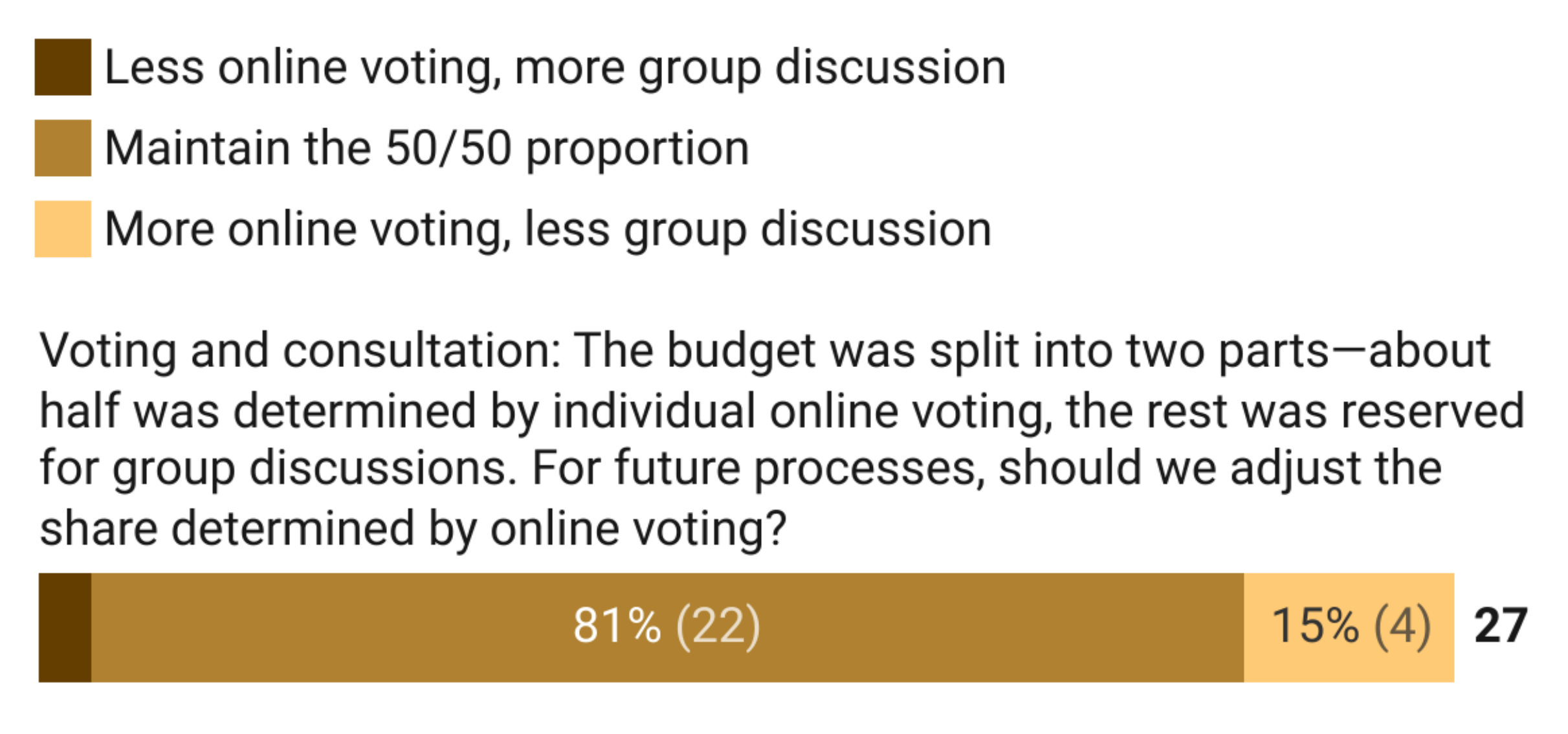}
\caption{Participant responses to how the ratio between voting, calculated using the MES algorithm, and deliberation should be adjusted in future processes.}
\label{fig:survey_portion}
\end{figure}
\begin{figure}[H]
\centering
\includegraphics[width=0.5\linewidth]{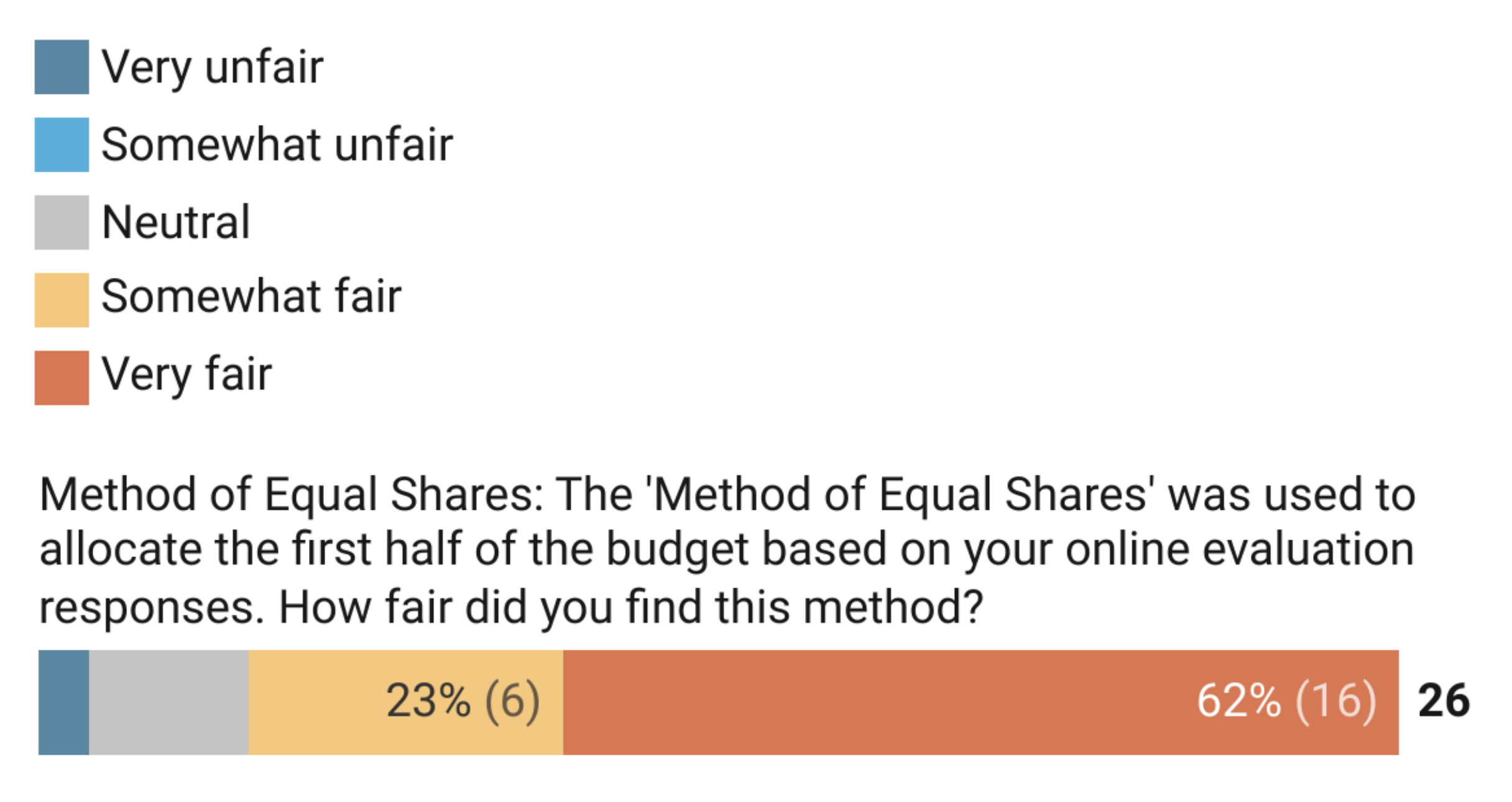}
\caption{Participant responses to the fairness of the Method of Equal Shares.}
\label{fig:survey_mes}

\end{figure}
\clearpage
\section{ReadTheRoom}

\begin{figure*}[h]
  \includegraphics[width=\linewidth]{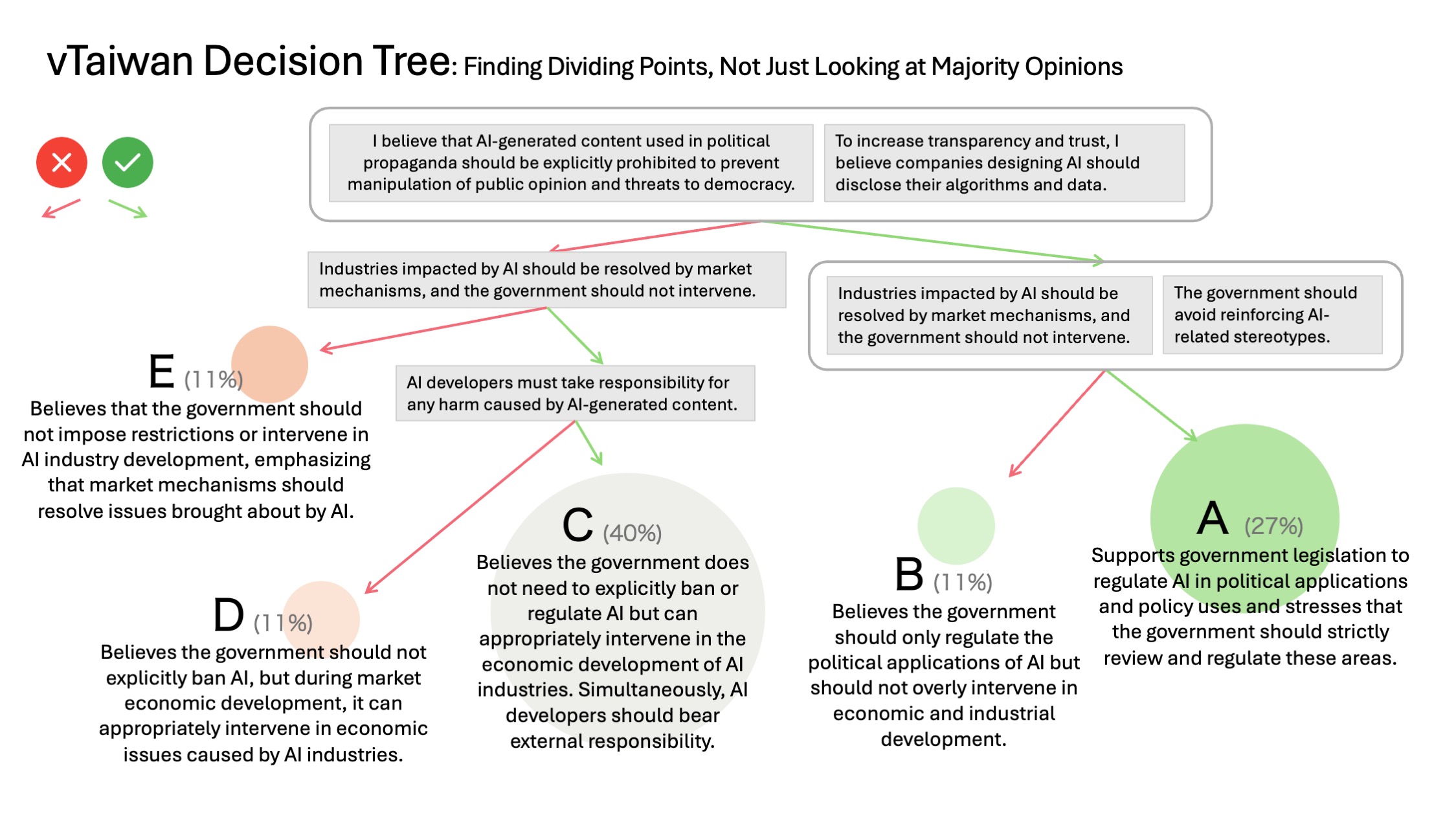}
  \caption{Translated Decision Tree Screenshot from the vTaiwan AI Regulation Deliberation Workshop on December 20, 2024. \small\normalfont Based on data collected through the Polis wiki-survey, this decision tree visualises how different opinion groups diverge in their opinions on AI governance. The opinion space is summarised using ChatGPT, interpreting key statements that stand out within each group. This illustration was used during the workshop to explain to the participants why certain statements were used as the central theme of the deliberative event.}
  \label{fig:tree}
\end{figure*}

\begin{figure*}[H]
\centering
\begin{minipage}[t]{0.49\linewidth}
    \centering
    \includegraphics[width=\textwidth]{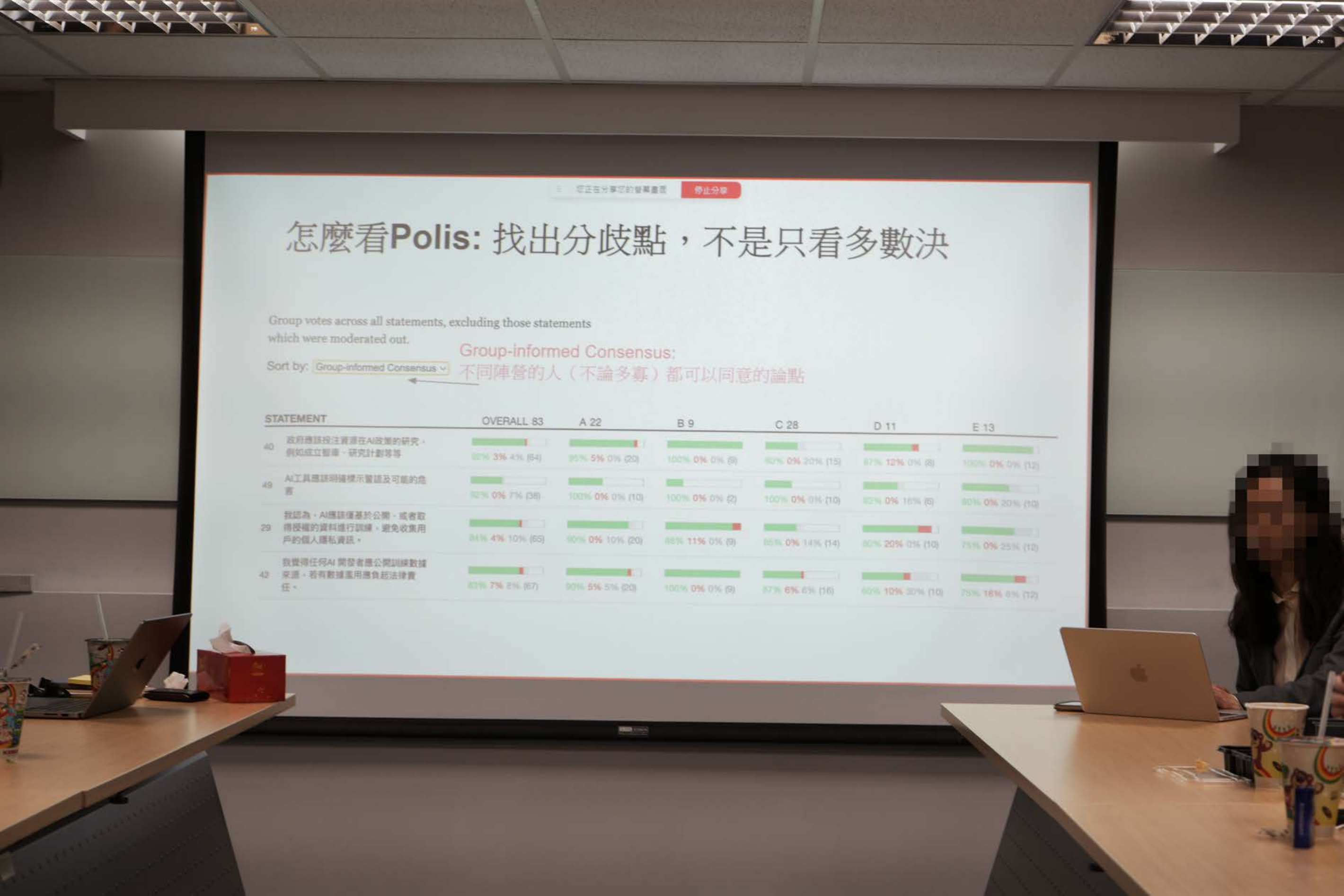}
    \caption{The vTaiwan Deliberation Workshop on AI Regulation held in Taipei, Taiwan, on December 20, 2024. The moderator presented the Polis report showcasing the \enquote{group informed consensus}, calculated using the online votes collected prior to the event. (Photo credit: vTaiwan)}
    \label{fig:polis}
\end{minipage}%
\hfill
\begin{minipage}[t]{0.49\linewidth}
    \centering
    \includegraphics[width=\textwidth]{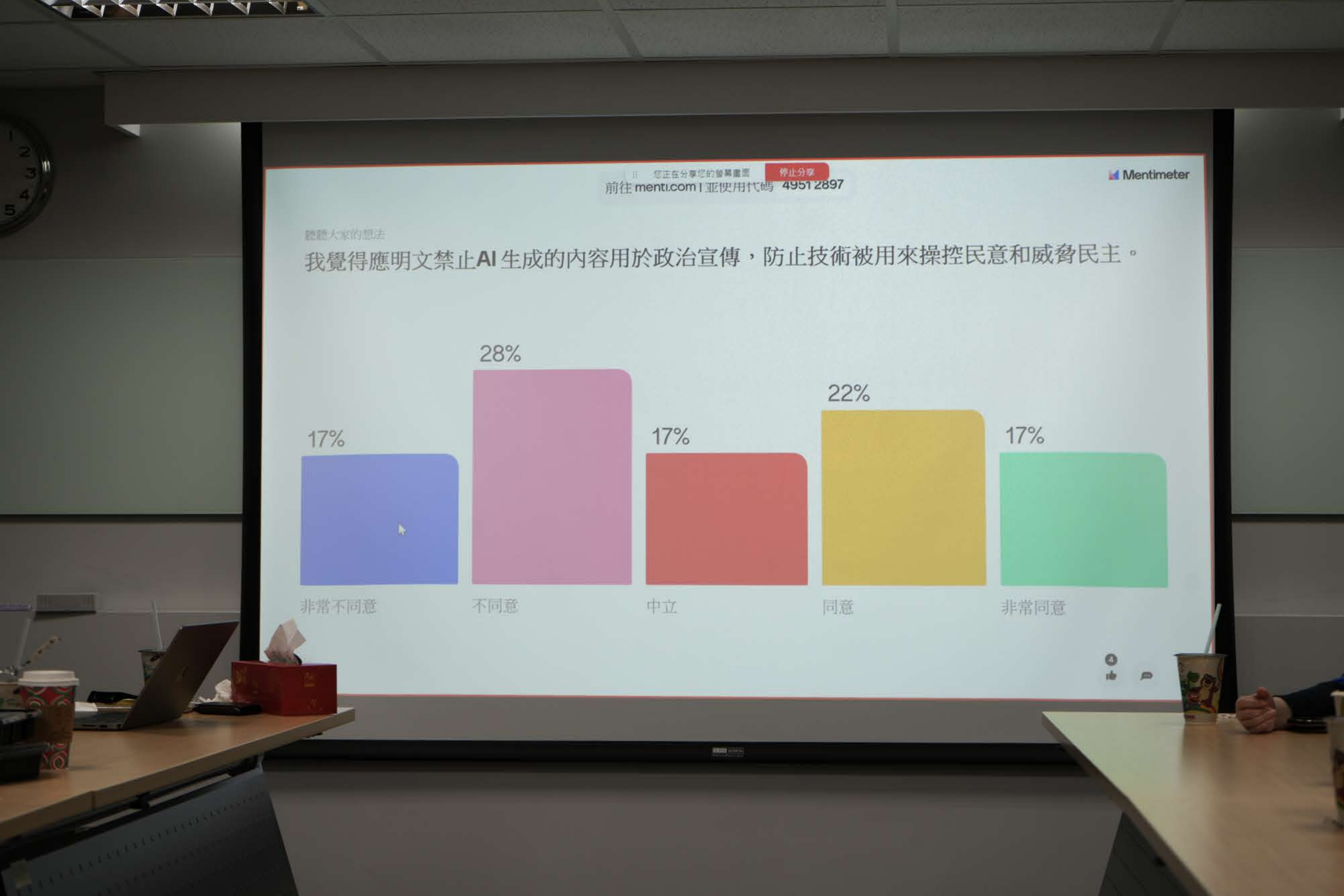}
    \caption{The vTaiwan ReadTheRoom game uses an interactive voting platform like Mentimeter to encourage participants to be open to different ideas. (Photo credit: vTaiwan)}
    \label{fig:readtheroom}
\end{minipage}
\end{figure*}
\begin{figure*}[h]
  \includegraphics[width=\linewidth]{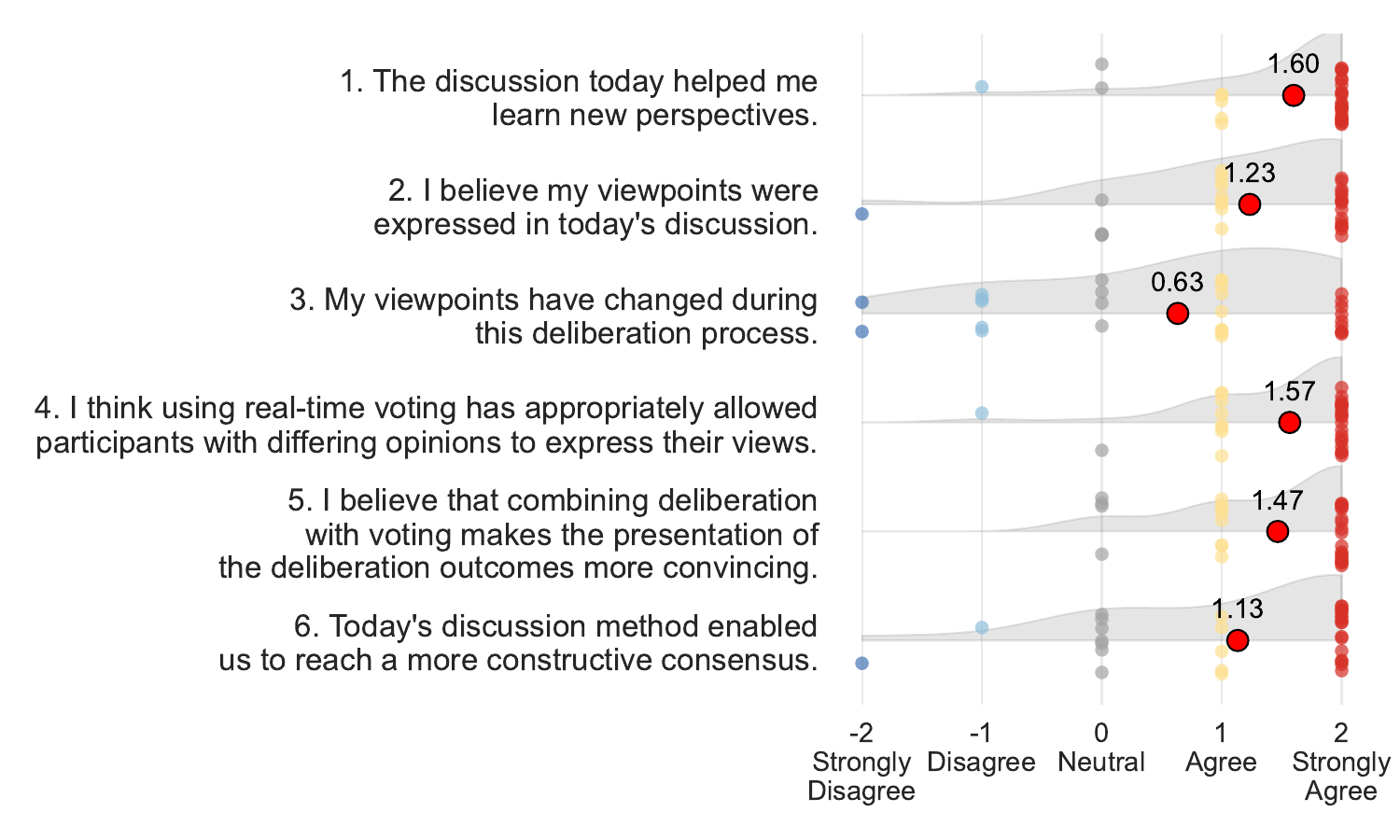}
  \caption{\textbf{Survey Responses from the vTaiwan Deliberation.} This diagram shows participant responses to six survey questions after deliberation. The vertical axis lists the questions, and the horizontal axis represents a Likert scale from -2 (\textit{Strongly Disagree}) to 2 (\textit{Strongly Agree}). Grey areas indicate the kde distribution of responses, coloured dots represent individual responses (blue for disagreement, grey for neutrality, orange to red for agreement), and red circles show the mean response with annotated values.}
  \label{fig:survey}
\end{figure*}

\end{document}